\documentclass[apj]{emulateapj}
\slugcomment{{\sc Accepted to ApJ:} June, 2013}
\usepackage{latexsym}
\usepackage{amsmath} 
\usepackage{amsfonts} 
\usepackage{amssymb}
\usepackage{makeidx} 
\usepackage{graphicx} 
\usepackage{natbib}
\usepackage{footnote}
\usepackage{color}

\usepackage[caption=false]{subfig}

\newcommand{\um}{\mbox{\,$\mu$m}}
\newcommand{\xxmark}{\text{\sffamily X}}

\begin{document} 
\title{Earliest Stages of Protocluster Formation: Substructure and Kinematics of Starless Cores in Orion} 
\author{Katherine I.\ Lee\altaffilmark{1}, Leslie W.\ Looney\altaffilmark{1,2}, Scott Schnee\altaffilmark{2}, Zhi-Yun Li\altaffilmark{3}} 
\altaffiltext{1}{Department of Astronomy, University of Illinois, Urbana, IL 61801, USA; ijlee9@illinois.edu}
\altaffiltext{2}{National Radio Astronomy Observatory, Charlottesville, VA 22903, USA}
\altaffiltext{3}{Department of Astronomy, University of Virginia, Charlottesville, VA 22904, USA}

\begin{abstract} 

We study the structure and kinematics of nine 0.1~pc-scale cores in Orion with the IRAM 30-m telescope and at higher
resolution eight of the cores with CARMA, using
CS(2-1) as the main tracer.
The single-dish moment zero maps of the starless cores show
single structures with central 
column densities ranging from 7 to $42 \times 10^{23}$ cm$^{-2}$ and 
LTE masses from 20 M$_{\sun}$ to 154 M$_{\sun}$.
However, at the higher CARMA resolution (5\arcsec), all of the cores
except one fragment into 3 - 5 components. The number of fragments is 
small compared to that found in some turbulent fragmentation 
models, although inclusion of magnetic fields may reduce the 
predicted fragment number and improve the model agreement.   
This result demonstrates that fragmentation from parsec-scale molecular clouds to sub-parsec cores continues
to take place inside the starless cores.  
The starless cores and their fragments are embedded in larger 
filamentary structures, which likely played a role in the core formation 
and fragmentation.
Most cores show clear velocity gradients, with magnitudes ranging 
from 1.7 to 14.3 km s$^{-1}$ pc$^{-1}$. We modeled one of them in
detail, and found that its spectra are best explained by a converging 
flow along a filament toward the core center; the gradients in other
cores may be modeled similarly. We infer a mass inflow
rate of $\sim 2\times 10^{-3}$ M$_{\odot}$~yr$^{-1}$, which is in
principle high enough to overcome radiation pressure and allow for
massive star formation. However, the core contains multiple fragments,
and it is unclear whether the rapid inflow would feed the growth of 
primarily a single massive star or a cluster of lower mass objects. 
We conclude that fast, supersonic converging flow along filaments
play an important role in massive star and cluster formation.

\end{abstract}

\section{Introduction}
\begin{figure*}
\begin{center}
\includegraphics[scale=0.5]{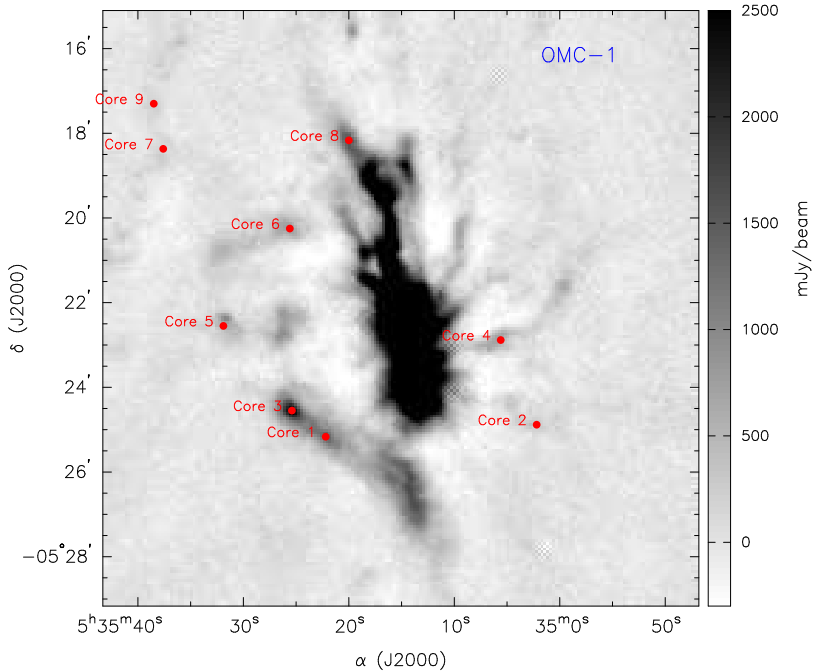}
\caption{The 850 $\um$ dust continuum image of Orion A-North from the JCMT SCUBA archive.
The positions of the nine cores detected with the IRAM 30-meter telescope are plotted with red circles.}
\label{fig:orion}
\end{center}
\end{figure*}

Star formation involves a complicated interplay between
turbulence, magnetic fields, and gravity. While the understanding 
of low-mass star formation has advanced over decades
\citep[e.g.,][]{2007ARA&A..45..565M}, that of massive star formation
has progressed more slowly. One difficulty is that massive protostars 
generate a much stronger radiation pressure that can strongly modify 
the gas accretion in the formation process 
\citep[see][]{2007ARA&A..45..481Z}.  Another is that massive stars 
appear to form in crowded environments of clusters \citep{2003ARA&A..41...57L}.
 This study aims to understand the conditions for massive star formation and their clustered environment 
at early stages.

There are two main competing scenarios in massive star formation.  
One of them is the ``turbulent core model" proposed by \citet{2003ApJ...585..850M}. 
In this model, the core is supported by supersonic turbulence and evolves on several free-fall timescale. 
The turbulent core is in quasi-static equilibrium and the formation of massive stars is a scaled-up version of low-mass star formation.
The final mass of a massive star is determined by the mass of its natal core and the stellar environment is unimportant.
The high pressure caused by supersonic turbulence results in high accretion rate ($> 10^{-3}$ M$_{\sun}$ yr$^{-1}$), overcoming 
the radiation pressure and continuing accretion process. 
In this model, the collapse is envisioned to be more or less 
monolithic with a relatively low level of fragmentation.

Alternatively, 
\citet{2004MNRAS.349..735B} and \citet{2006MNRAS.370..488B} proposed that massive star formation is a dynamical process that involves competitive accretion.
In a core that is dominated by supersonic turbulent motions, 
significant density fluctuations are generated due to turbulent support with gravity taking over in the densest regions.
Stellar seeds are created through this ``turbulent
  fragmentation" and eventually lead to clusters.
Massive stars form from the seeds located near the center of the
  cluster where the gravitational potential is deepest; these seeds 
win the competition for the reservoir of gas to grow to the highest masses.  
In this scenario, the final mass of a massive star is strongly influenced by their environment and has little correlation with the initial mass of the natal core.

A major difference of the two scenarios, which can be tested observationally, is the level of fragmentation. 
The turbulent core scenario envisions the existence of one massive starless core,
while the competitive accretion scenario requires a higher level of fragmentation in massive starless cores.
Therefore, our goal is to study the fragmentation in starless cores, the earliest stage of star formation where the initial conditions for massive star formation are still kept, and to provide insight to the formation of massive stars.

Due to angular resolution limitations, only recently has the study of fragmentation in starless/prestellar cores made significant progress.
Lately, a number of studies toward massive star-forming regions with
high angular resolutions using interferometers have begun to reveal
fragmentation in massive starless cores on 0.1 pc scale \citep[][]{2010A&A...524A..18B,2013ApJ...762..120P}.
These studies mostly focus on the massive star-forming sites associated with Infrared Dark Clouds (IRDCs), which are typically at a distance of few kilo-parsecs.
We chose Orion as our target region.  
At a distance of 414 pc \citep{2007A&A...474..515M}, 
Orion is the closest active star-forming region that contains massive
stars
\citep[e.g.,][]{1997AJ....113.1733H,1998ApJ...492..540H,1999ApJ...510L..49J}
as well as massive starless cores
\citep[e.g.,][]{nutter07,2008ApJS..175..277D,2010ApJ...710.1247S}.  
It provides an excellent opportunity to study the initial conditions
for massive star formation.

\section{Observations and Data Reduction}
\label{sect:obs}

\begin{deluxetable}{cccr}
\tablecaption{Coordinates of the IRAM 30-m detected sources}
\tabletypesize{\footnotesize}
\tablewidth{240pt}
\tablecolumns{4}
\tablehead{
\colhead{Source} & \colhead{R.A.(2000)} & \colhead{Dec.(2000)} & \colhead{850 $\um$ dust mass\tablenotemark{a}}
}
\startdata
Core 1 & 05:35:22.2 & -05:25:10 & 47.4 M$_{\sun}$ \\
Core 2 & 05:35:02.2 & -05:24:53 & 1.4 M$_{\sun}$ \\
Core 3 & 05:35:25.4 & -05:24:33 & 54.6 M$_{\sun}$ \\
Core 4 & 05:35:05.6 & -05:22:53 & 8.5 M$_{\sun}$ \\
Core 5 & 05:35:31.9 & -05:22:33 & 5.3 M$_{\sun}$ \\
Core 6 & 05:35:25.6 & -05:20:15 & 14.0 M$_{\sun}$ \\
Core 7 & 05:35:37.6 & -05:18:22 & 5.8 M$_{\sun}$ \\
Core 8 & 05:35:20.0 & -05:18:10 & 12.5 M$_{\sun}$ \\
Core 9 & 05:35:38.5 & -05:17:18 & 3.0 M$_{\sun}$
\enddata
\label{tbl:radec}
\tablenotetext{a}{The 850 $\um$ dust masses were calculated assuming a temperature of 20 K, a mass opacity of 0.01 cm$^{2}$ g$^{-1}$,
and a distance of 414 pc to Orion.  See \citet{nutter07} for more details.}
\end{deluxetable}

To fully investigate the substructure and kinematics of starless cores, we observed with
the IRAM 30-m single dish telescope and the the Combined Array for Research in Millimeter-wavelength (CARMA) interferometer.
Although we observed multiple molecular line tracers, CS(2-1) is mainly used for the analysis.
CS(2-1) is used extensively as a tracer for high-density gas.  
Although previous studies indicated that C-bearing species may be depleted during the prestellar phase \citep[e.g.,][]{1998A&A...336..309T},
several studies have also suggested that associated C-bearing molecules have not been frozen out at the very early stage of star formation.

\begin{deluxetable}{ccc}
\tablecaption{Coordinates of Non-detected Sources}
\tabletypesize{\footnotesize}
\tablewidth{240pt}
\tablecolumns{3}
\tablehead{
\colhead{Source Name\tablenotemark{a}} & \colhead{R.A.(2000)} & \colhead{Dec.(2000)}
}
\startdata
OrionAN-535031-52140 & 05:35:03.1 & -05:21:40 \\
OrionAN-535354-52130 & 05:35:35.4 & -05:21:30 \\
OrionAN-535181-52129 & 05:35:18.1 & -05:21:29 \\
OrionAN-535040-52046 & 05:35:04.0 & -05:20:46 \\
OrionAN-535146-51847 & 05:35:14.6 & -05:18:47 \\
OrionAN-535376-51822 & 05:35:37.6 & -05:18:22 \\
OrionAN-535200-51810 & 05:35:20.0 & -05:18:10 \\
OrionAN-535385-51718 & 05:35:38.5 & -05:17:18 \\
OrionAN-535217-51312 & 05:35:21.7 & -05:13:12 \\
OrionAN-535255-50237 & 05:35:25.5 & -05:02:37 \\
OrionAN-535207-50053 & 05:35:20.7 & -05:00:53
\enddata
\label{tbl:off}
\tablenotetext{a}{The source names are quoted from \citet{nutter07}.}
\end{deluxetable}

\begin{deluxetable*}{llrrrrrrrrr}
\tablecaption{Molecular Line Observations with the IRAM 30-m Telescope}
\tabletypesize{\footnotesize}
\tablewidth{0pt}
\tablecolumns{11}
\tablehead{
\colhead{Line transition} & \colhead{Frequency} & \multicolumn{9}{c}{Signal-to-Noise Ratio (from the peak intensity)} \\
\cline{3-11} \\
\colhead{} & \colhead{(GHz)} & \colhead{Core 1} & \colhead{Core 2} & \colhead{Core 3} & \colhead{Core 4} & \colhead{Core 5} & \colhead{Core 6} & \colhead{Core 7} & \colhead{Core 8} & \colhead{Core 9} \\
}
\startdata
H$^{13}$CO$^{+}$(1-0) & 86.754330 & \xxmark & \xxmark & \xxmark & \nodata & \xxmark & \nodata & \nodata & \nodata & \nodata \\
C$^{34}$S(2-1) & 96.412982 & 8.0 & 5.3 & 13.4 & \nodata & 4.7 & \nodata & \xxmark & 10.6 & \xxmark \\
CS(2-1) & 97.980968 & 28.0 & 13.1 & 43.3 & 96.2 & 37.0 & 5.6 & 70.2 & 137.9 & 28.7 \\
$^{18}$CO(1-0) & 109.782182 & 7.7 & 6.5 & 10.9 & \nodata & \nodata & \nodata & \nodata & 18.5 & \nodata \\
CS(3-2) & 146.969049 & 20.0 & 8.0 & 36.0 & \nodata & 12.3 & \nodata & 7.1 & 20.8 & \xxmark \\
N$_{2}$D$^{+}$(2-1) & 154.217206 & \xxmark & \xxmark & \xxmark & \nodata & \xxmark & \nodata & \nodata & \nodata & \nodata \\
$^{18}$CO(2-1) & 219.560319 & 5.7 & 3.3 & 9.8 & \nodata & \nodata & \nodata & \nodata & 6.8 & \nodata \\
$^{12}$CO(2-1) & 230.537990 & 13.7 & \xxmark & 11.5 & 20.8 & 4.7 & 52.3 & 55.6 & 92.8 & 50.0
\enddata
\label{tbl:detection}
\tablecomments{\xxmark \ means no detection (below the 3$\sigma$ level) with the molecular line.  \nodata means the molecular line was not observed.}
\end{deluxetable*}

\begin{deluxetable*}{rrrrr}
\tablecaption{Sensitivity Limits from the IRAM and CARMA Observations}
\tabletypesize{\footnotesize}
\tablewidth{0pt}
\tablecolumns{5}
\tablehead{
\colhead{Source} & \colhead{IRAM CS(2-1)} & \colhead{IRAM CS(3-2)} & \colhead{IRAM C$^{34}$S(2-1)} & \colhead{CARMA CS(2-1)}
}
\startdata
Core 1 & 0.24 K & 0.67 K & 0.18 K & 0.15 Jy beam$^{-1}$ \\
Core 2 & 0.15 K & 0.66 K & 0.34 K & 0.11 Jy beam$^{-1}$ \\
Core 3 & 0.10 K & 0.54 K & 0.23 K & 0.20 Jy beam$^{-1}$ \\
Core 4 & 0.085 K & \nodata & \nodata & 0.20 Jy beam$^{-1}$ \\
Core 5 & 0.10 K & 0.41 K & 0.18 K & 0.11 Jy beam$^{-1}$ \\
Core 6 & 0.09 K & \nodata & \nodata & 0.15 Jy beam$^{-1}$ \\
Core 7 & 0.075 K & \nodata & \nodata & 0.16 Jy beam$^{-1}$ \\
Core 8 & 0.1 K   & 0.75 K  & 0.32 K  & 0.18 Jy beam$^{-1}$ \\
Core 9 & 0.075 K & \nodata & \nodata & \nodata
\enddata
\label{tbl:sens}
\end{deluxetable*}

Our sample of starless cores were chosen from \citet{nutter07}. 
They conducted a large survey in Orion at 850 $\um$ dust continuum with the Submillimetre Common User Bolometer Array (SCUBA).
By comparing the survey with the \textit{Spitzer IRAC} catalog, 
the study provided a complete catalog of prestellar cores and protostellar cores down to the lower completeness limit at $\sim$ 0.3 M$_{\sun}$ in the Orion A North, A South, Orion B North and B South regions.
We chose 16 prestellar cores from the catalog with dust masses\footnotemark[1] ranging from 1 M$_{\sun}$ to 50 M$_{\sun}$
and observed them with the IRAM 30-m telescope.
The reason to include a few cores with low dust mass estimates was to
compare them with massive cores; 
surprisingly, some cores with low dust mass estimates turned out have large gas mass estimates from our observations (see Sect.\ \ref{sect:column}).
Of the 16 sources in the IRAM 30-m sample, 9 were detected in at least one molecular line tracer.
Figure \ref{fig:orion} shows the positions of the 9 detected cores, and
Table \ref{tbl:radec} lists the coordinates and the masses calculated from the 850 $\um$ dust continuum observations of those cores.
Table \ref{tbl:off} lists the coordinates of the non-detected sources.  
The CARMA observations were performed toward 8 of the 9 starless cores that had been detected with the IRAM observations.

\subsection{The sample}
\label{sect:sample}

The cores are mostly located in the filamentary structures around the periphery of the central Orion Molecular Cloud 1 (OMC-1).
Core 1 and Core 3 are in the Orion Bar photon-dominated (PDR) region \citep[e.g.,][]{2003ApJ...597L.145L}.
Core 2 and Core 4 are associated with the ``radiating filaments" \citep{1999ApJ...510L..49J} from OMC-1 of which the formation mechanisms are still unclear \citep{2009ApJ...700.1609M}.
In particular, Core 4 is close to the Orion BN/KL region which is observed with powerful outflows and ``H$_{2}$ fingers" \citep{2009ApJ...704L..45Z,2012A&A...544L..19P}; 
however, the location of Core 4 is in a larger spatial scale structure and is not associated with the HH bullets or H$_{2}$ fingers \citep{2012MNRAS.422..521B}.
Core 7 and Core 9 are associated with a filament north-east to OMC-1, which has been suggested as a PDR (Shimajiri et al.\ 2013, in prep).   
 
\footnotetext[1]{The term ``dust mass" used in this paper refers to a total mass from dust and gas derived from dust emission by assuming a gas-to-dust ratio (typically 100).}

\subsection{IRAM 30-m observations}
\label{sect:obsiram}

The observations were performed in August 2010 toward the sixteen starless cores in Orion.
The heterodyne receiver EMIR was used.
The bands E090, E150 and E230 in combination captured various lines.
The bands E090 and E150 were used to perform the molecular line observations with CS(2-1) at 97.980968 GHz, C$^{34}$S(2-1) at 96.41298 GHz and CS(3-2) at 146.96905 GHz.
We used VESPA as the spectral back-end with a spectral resolution of 40 kHz ($\sim$ 0.06 km s$^{-1}$ at 3 mm) and a total bandwidth of 80 MHz.
On-the-fly mapping was performed with both horizontal and vertical scanning to span an area of about 2\arcmin \ by 2\arcmin \ for each source.
The observations were performed in position-switching mode with off-position at $05^{h}36^{m}15.0^{s}$, $-05\deg02\arcmin34\arcsec$ \citep{2007ApJ...665.1194I}.
Calibration scans were taken about every fifteen minutes.
The pointing was checked every two hours.  
The beam size (full-width half-power) is $\sim$25.5\arcsec at the frequency of CS(2-1) and C$^{34}$S(2-1), and 17\arcsec \ at the frequency of CS(3-2).
The beam efficiency\footnotemark[2]\ (Beff) is 81\% for CS(2-1) and C$^{34}$S(2-1), and 74\% for CS(3-2).

The data reduction was done with the CLASS package from the GILDAS\footnotemark[3] software. 
All the data are re-gridded to have at least three pixels in one beam size.
All sixteen sources were observed with CS(2-1) and $^{12}$CO(2-1); only nine of which showed detections (Table \ref{tbl:detection}).  

\footnotetext[2]{See http://www.iram.es/IRAMES/mainWiki/Iram30mEfficiencies}
\footnotetext[3]{See http://www.iram.fr/IRAMFR/GILDAS/}

\begin{deluxetable}{cccc}
\tablecaption{Velocity ranges for the IRAM 0$^{th}$ moment maps}
\tabletypesize{\footnotesize}
\tablewidth{240pt}
\tablecolumns{4}
\tablehead{
\colhead{Source} & \colhead{CS(2-1)} & \colhead{CS(3-2)} & \colhead{C$^{34}$S(2-1)} \\
\colhead{} & \colhead{(km s$^{-1}$)} & \colhead{(km s$^{-1}$)} & \colhead{(km s$^{-1}$)}
}
\startdata
Core 1 & 11.91 - 9.16 & 12.35 - 8.88 & 11.04 - 9.44 \\
Core 2 & 13.23 - 6.77 & 12.23 - 6.93 & 11.35 - 8.84 \\
Core 3 & 11.73 - 8.57 & 12.23 - 8.25 & 11.04 - 9.24 \\
Core 4 & 12.03 - 5.58 & \nodata & \nodata \\
Core 5 & 8.57 - 4.38 & 8.25 - 4.86 & 8.45 - 7.45 \\
Core 6 & 8.39 - 5.04 & \nodata & \nodata \\
Core 7 & 11.91 - 6.77 & \nodata & \nodata \\
Core 8 & 11.91 - 7.49 & 12.00 - 6.83 & 12.24 - 8.45 \\
Core 9 & 12.15 - 7.37 & \nodata & \nodata
\enddata
\label{tbl:vrange}
\end{deluxetable}

\subsection{CARMA observations}
CARMA is a heterogeneous array combining three types of antenna: six 10-meter antennas, nine 6-meter antennas and eight 3.5-meter antennas. 
The data presented in this paper used the cross-correlated data from the 10-meter antennas and the 6-meter antennas.
The CARMA observations were performed toward 8 starless cores (from the 9 sources detected by the IRAM 30-m telescope) between May 2010 and November 2011. 
Only one source (core 8) out of the eight observed sources was
observed with both the D and E array configuration, and the 
remaining seven cores were observed with only the D array.
The data presented in this paper focus on CS(2-1) at 97.98096 GHz, with one band for N$_{2}$H$^{+}$(1-0) at 93.17383 GHz and one band for the continuum observation.
The projected baselines of the D array range from 11 m to 150 m, providing sensitivity to spatial scales up to $\sim$ 30\arcsec \ and a synthesized beam of $\sim \ 5\arcsec$ at 3 mm. 
The E array has the projected baselines ranging from 8 m to 66 m, providing sensitivity to spatial scales up to $\sim$ 40\arcsec \ and a synthesized beam of $\sim \ 7\arcsec$ at 3 mm.
The spectral resolutions are 0.15 km s$^{-1}$ for core 1, 2, 4, 5 and 0.06 km s$^{-1}$ for core 3, 6, 7, 8. 
The amplitude calibration is estimated to be 10\%, and the uncertainties discussed afterwards are only statistical, not systematic.
All the data reduction were done with the MIRIAD software \citep{1995ASPC...77..433S}.

The sensitivity limits of all the data present from IRAM and CARMA in this paper are summarized in Table \ref{tbl:sens}.

\subsection{Herschel and JCMT archival data}

For the 8 sources observed by CARMA, we also present the 500 $\micron$ images
from the \textit{Herschel} archival data as well as the  850 $\um$ image from the JCMT archival data. 
The \textit{Herschel} 500 $\um$ image is downloaded from the Herschel Science Archive\footnotemark[4] (HSA).
The data were taken in September, 2009 with the Spectral and Photometric Imaging Receiver (SPIRE).
The observation identifier number is 1342184386, and the data presented in this paper is calibrated to level 2.  
The 850 $\um$ image from JCMT SCUBA-2 is downloaded from the site of public processed data in the JCMT science archive\footnotemark[5].
The project number is M09BI121 and the data was taken in Feb, 2010.

\footnotetext[4]{See http://herschel.esac.esa.int/Science\_Archive.shtml}
\footnotetext[5]{See http://www.cadc.hia.nrc.gc.ca/jcmt/search/product/}

\section{Results and Data Analysis I: Morphology and Properties}
\subsection{IRAM maps}
\subsubsection{CS(2-1), CS(3-2) and C$^{34}$S}
\label{sect:iram}

\begin{figure*}
\begin{center}
\includegraphics[scale=0.7,angle=0]{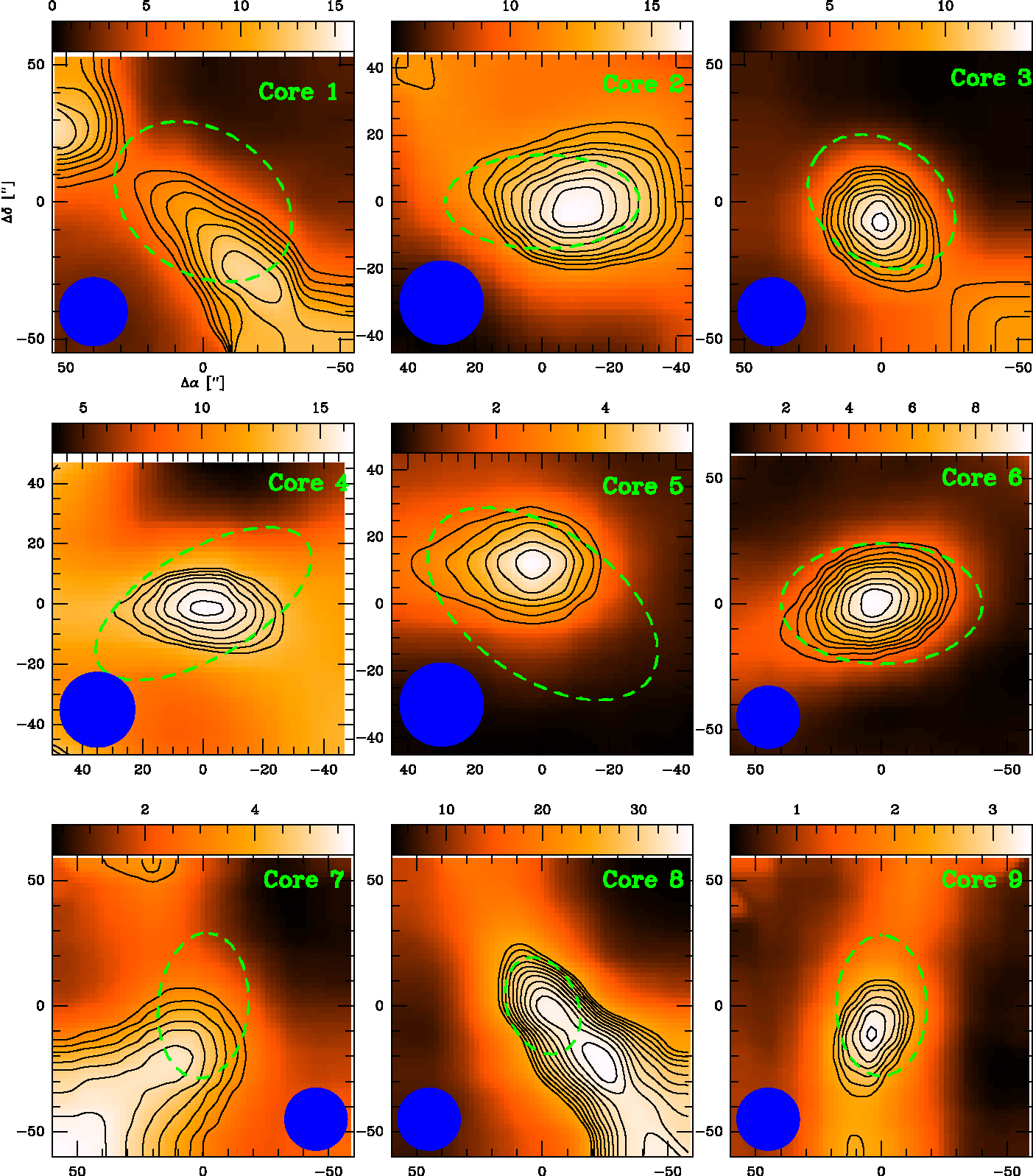}
\caption{The $0^{th}$ moment maps from the IRAM CS(2-1) observations toward the nine starless cores.  The blue circle in each panel indicates the beam size (25\arcsec) of the IRAM 30-m telescope.
The green dashed ellipses are the definitions of 850 $\um$ dust cores from \citet{nutter07}.
Most of the cores show single peaks and are associated with elongated structures.
The contour levels ($\sigma$, starting level, interval) are as following.
Core 1: $\sigma$=0.1 K km s$^{-1}$, 70$\sigma$, 10$\sigma$.
Core 2: $\sigma$=0.1 K km s$^{-1}$, 120$\sigma$, 5$\sigma$.
Core 3: $\sigma$=0.04 K km s$^{-1}$, 175$\sigma$, 20$\sigma$.
Core 4: $\sigma$=0.05 K km s$^{-1}$, 265$\sigma$, 10$\sigma$.
Core 5: $\sigma$=0.05 K km s$^{-1}$, 55$\sigma$, 10$\sigma$.
Core 6: $\sigma$=0.04 K km s$^{-1}$, 100$\sigma$, 15$\sigma$.
Core 7: $\sigma$=0.04 K km s$^{-1}$, 80$\sigma$, 10$\sigma$.
Core 8: $\sigma$=0.05 K km s$^{-1}$, 500$\sigma$, 20$\sigma$.
Core 9: $\sigma$=0.04 K km s$^{-1}$, 60$\sigma$, 4$\sigma$.
}
\label{fig:850dust}
\end{center}
\end{figure*}

\begin{deluxetable*}{llcccccccc}
\tablecaption{Properties of the five Cores with CS(2-1) and C$^{34}$S(2-1) IRAM 30-m Detections}
\tabletypesize{8pt}
\tablewidth{0pt}
\tablecolumns{10}
\tablehead{
\colhead{Source} & \colhead{Transition} & \colhead{T$_{A}$} & \colhead{$\tau$} & \colhead{$\int T_{A} dv$} & \colhead{T$_{ex,CS(2-1)}$} & \colhead{N$_{H_{2}}$} & \colhead{$\Sigma_{H_{2}}$} & \colhead{M$_{H_{2}}$} & \colhead{n$_{H_{2}}$} \\
\colhead{} & \colhead{} & \colhead{(K)} & \colhead{} & \colhead{(K km s$^{-1}$)} & \colhead{(K)} & \colhead{($\times 10^{23}$ cm$^{-2}$)} & \colhead{(g cm$^{-2}$)} & \colhead{(M$_{\sun}$)} & \colhead{($\times 10^{6}$ cm$^{-3}$)}
}
\startdata
Core 1  & CS(2-1)               & 6.81  & 2.24  & 12.46 $\pm$ 0.1 & 12.76 & 7.28 $\pm$0.06 & 1.81 $\pm$ 0.01 & 26.60 $\pm$ 0.15 & 5.44 $\pm$ 0.03 \\
        & C$^{34}$S(2-1)        & 0.72  & 0.10  &                 &       &                &                 &                  &                 \\
Core 2  & CS(2-1)               & 3.90  & 2.69  & 15.74 $\pm$ 0.1 & 8.38  & 10.02 $\pm$ 0.06 & 2.48 $\pm$ 0.01 & 36.60 $\pm$ 0.15 & 7.47 $\pm$ 0.03 \\
        & C$^{34}$S(2-1)        & 0.47  & 0.12  &                 &       &                  &                  &                 &                 \\
Core 3  & CS(2-1)               & 5.32  & 3.66  & 10.67 $\pm$ 0.04 & 10.02 & 8.85 $\pm$ 0.03 & 2.19 $\pm$ 0.01 & 32.34 $\pm$ 0.15 & 6.61 $\pm$ 0.03 \\
        & C$^{34}$S(2-1)        & 0.82  & 0.16  &                  &       &                 &                 &                  &                 \\
Core 5  & CS(2-1)               & 2.64  & 4.78  & 4.63 $\pm$ 0.05 & 6.42  & 5.32 $\pm$ 0.06 & 1.32 $\pm$ 0.01 & 19.40 $\pm$ 0.15 & 3.97 $\pm$ 0.03 \\
        & C$^{34}$S(2-1)        & 0.51  & 0.21  &                 &       &                 &                 &                  &                 \\
Core 8  & CS(2-1)               & 12.47 & 4.69  & 31.88 $\pm$ 0.05 & 19.01 & 42.14 $\pm$ 0.07 & 10.45 $\pm$ 0.02 & 154.00 $\pm$ 0.3 & 31.46 $\pm$ 0.06 \\
        & C$^{34}$S(2-1)        & 2.37  & 0.21  &                  &       &                  &                  &                  &

\enddata
\label{tbl:coldensity}
\end{deluxetable*}

We detected 9 out of the 16 cores in the IRAM 30-m sample.  We
assessed the \textit{Spitzer} 8.0 $\um$ maps and the IRAM CO data
(C$^{12}$O(2-1), C$^{18}$(2-1)) and confirmed that these are indeed
starless cores that lack detected infrared counterparts and outflows.
To determine the spatial extent of each core, we examined the data cube and
selected the velocity range to make 0$^{th}$ moment maps based on
3$\sigma$ levels in channel maps.  We then defined a projected core as the lowest closed contour
in 0$^{th}$ moment maps over that range (Figure \ref{fig:850dust})
and masked based on that contour.  
The velocity
ranges used for 0$^{th}$ moment maps of CS(2-1), CS(3-2), and
C$^{34}$S(2-1) are summarized in Table \ref{tbl:vrange}; the range is the 
3$\sigma$ detection range of molecular emission in the defined core.
The defined cores have high signal-to-noise\footnotemark[6] in the 0th moment maps, implying that the cores are likely surrounded by large-scale emission.  While this large-scale emission will have some effect on our core properties (e.g., fluxes, masses), it is difficult to disentangle the contributions.

\footnotetext[6]{The noises on 0$^{th}$ moment maps were calculated with $\sigma_{I} = \sigma_{T} N_{ch}^{1/2} \Delta v$, where $\sigma_{I}$ is the noise on 0$^{th}$ moment maps, $\sigma_{T}$ is the noise on channel maps, $N_{ch}^{1/2}$ is the number of channels in the summed velocity range, and $\Delta v$ is the channel width.}

We note that Core 1, 7, and 8 are associated with ambient extended structures
and no closed contours can be determined.  For these three cores, masking
is performed based on the lowest contour that shows distinct
structures different than the extended structures.  Although in general the morphology
and sizes of our gas derived cores are consistent with the 850 $\um$
dust continuum emission derived cores \citep{nutter07}, there are
some clear discrepancies (e.g. Cores 5 and 7) that could suggest
that the dust and gas traced by CS(2-1) are not well correlated in
the early stage of star formation, which is also seen in other
studies \citep{morata2012}.

The IRAM CS(2-1) sources are mostly single-peaked (Figure \ref{fig:850dust}), 
and are usually associated with non-spherical, elongated structures. 
The non-spherical morphologies agree with previous studies of dense cores, which showed that the majority of dense cores are tri-axial including prolate or oblate \citep[e.g.,][]{2007MNRAS.379L..50T}. 
For 5 of the cores, we can compare the CS(3-2) and C$^{34}$S(2-1) emission with the CS(2-1) emission (Figure \ref{fig:cs}), 
and the 0$^{th}$ moment images are mapped using the full range of the lines.  
CS(3-2) and C$^{34}$S(2-1) are in general optically thinner than CS(2-1). 
These three lines show consistent morphologies, and the CS(3-2) peaks well coincide with the CS(2-1) peaks.
There are a few differences; for example, 
the C$^{34}$S(2-1) emission shows more than one peak for core 2, but shows only one peak for core 8.
Both CS(3-2) and C$^{34}$S show more spherical shape of the core than the CS(2-1) emission.

H$^{13}$CO$^{+}$(1-0) and N$_{2}$D$^{+}$(2-1) are not detected toward any core (Table \ref{tbl:detection}).
A classification between ``early-time" and ``late-time" molecules have been suggested by several studies based on the time at which these molecules reached their peak abundance \citep[e.g.,][]{1998A&A...336..309T,morata2003}.
CS(2-1) is in general classified as an ``early-time" tracer while H$^{13}$CO$^{+}$(1-0) and nitrogen-bearing species are ``late-time" tracers \citep[e.g.,][]{morata2005}.
Therefore, the detection of CS(2-1) and the non-detections of H$^{13}$CO$^{+}$(1-0) and N$_{2}$D$^{+}$(2-1) in Cores 1, 2, 3, 5 suggest that the cores are in the very early stage of the evolution
and are chemically young, before the depletion of CS(2-1) becomes significant \citep[e.g.,][]{2002ApJ...569..815T}.
However, the detail of the chemistry depends on various models \citep[e.g.,][]{2012ApJ...751..105V}.

\begin{figure*}
\begin{center}
\includegraphics[scale=0.7,angle=0]{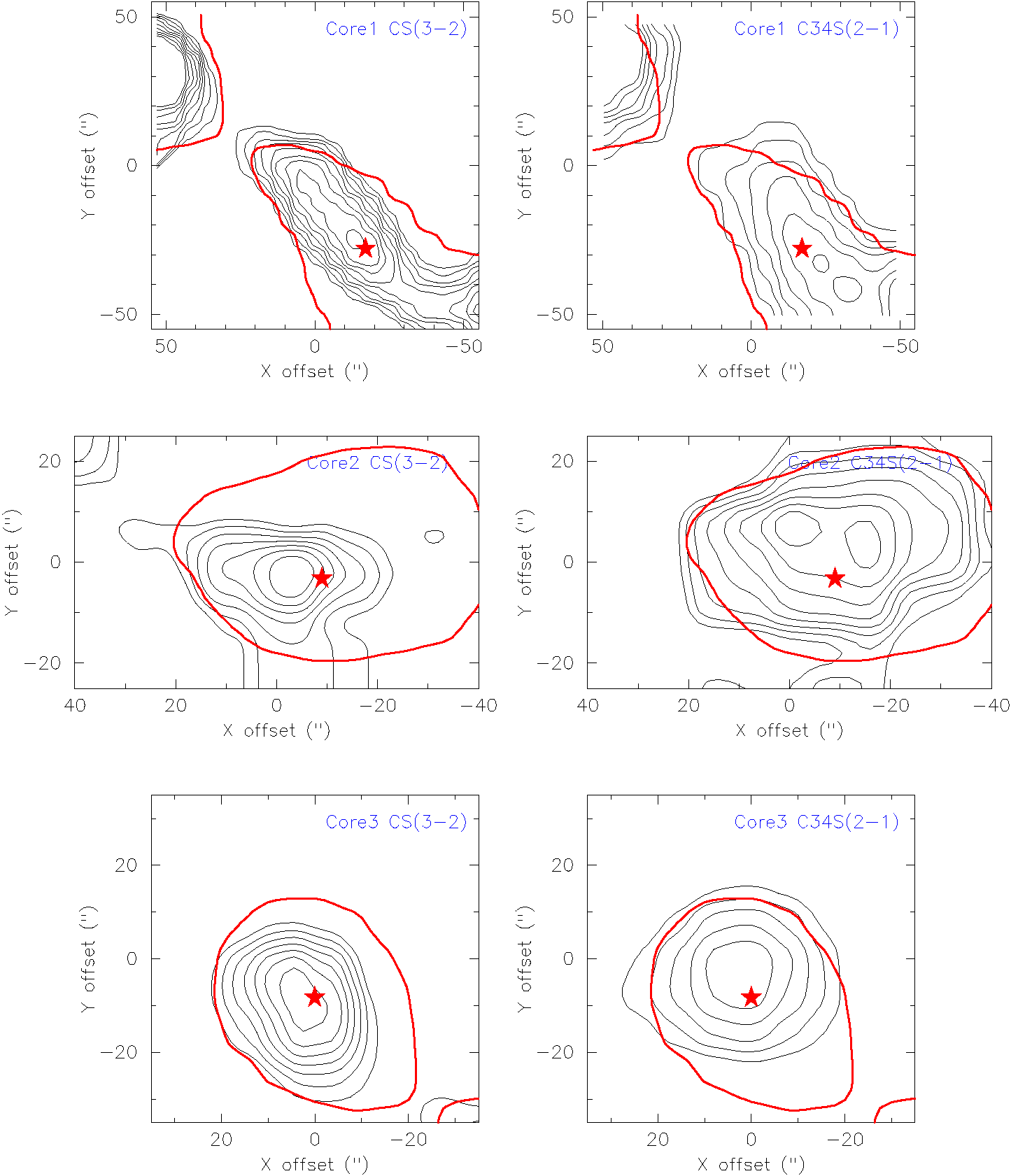}
\caption{$0^{th}$ moment maps from the IRAM CS(3-2) and C$^{34}$S(2-1) observations for the five starless cores.  The black contours in the left panels show CS(3-2) and the black contours in the right panels show C$^{34}$S(2-1).  The red contours are the IRAM CS(2-1) emission (lowest contours in Fig.\ \ref{fig:850dust}) and the red stars are where the CS(2-1) emission peak.
The contour levels ($\sigma$, starting level, interval) for CS(3-2) are:
Core 1: $\sigma=0.25$ K km s$^{-1}$, 40$\sigma$, 5$\sigma$; 
Core 2: $\sigma=0.30$ K km s$^{-1}$, 34$\sigma$, 2$\sigma$;
Core 3: $\sigma=0.22$ K km s$^{-1}$, 50$\sigma$, 10$\sigma$;
Core 5: $\sigma=0.15$ K km s$^{-1}$, 25$\sigma$, 5$\sigma$;
Core 8: $\sigma=0.34$ K km s$^{-1}$, 50$\sigma$, 10$\sigma$.
The contour levels ($\sigma$, starting level, interval) for C$^{34}$S(2-1) are:
Core 1: $\sigma=0.07$ K km s$^{-1}$, 8$\sigma$, 2$\sigma$;
Core 2: $\sigma=0.16$ K km s$^{-1}$, 8$\sigma$, 0.5$\sigma$;      
Core 3: $\sigma=0.1$ K km s$^{-1}$, 10$\sigma$, 2$\sigma$;
Core 5: $\sigma=0.06$ K km s$^{-1}$, 5$\sigma$, 2$\sigma$;
Core 8: $\sigma=0.16$ K km s$^{-1}$, 15$\sigma$, 5$\sigma$.
}
\label{fig:cs}
\end{center}
\end{figure*}

\subsubsection{CS(2-1) column densities}
\label{sect:column}

As we have IRAM 30-m CS(2-1) and C$^{34}$S(2-1) emission for 5 of the cores, 
we are able to estimate their optical depths and peak column densities.
We first calculate the optical depths of CS(2-1) and C$^{34}$S(2-1) by
\begin{equation}
\frac{T_{MB}(C^{34}S)}{T_{MB}(CS)} = \frac{1-exp(-\tau)}{1-exp(-\tau f)},
\end{equation}
where T$_{MB}$ is the main beam temperature of the molecular line, and f is the CS to C$^{34}$S ratio. 
T$_{MB}$ is calculated from the observed antenna temperature divided by the main beam efficiency (0.8 as reported in Sect.\ \ref{sect:obsiram}).
T$_{MB}$(CS) and T$_{MB}$(C$^{34}$S) are measured from the flux within the beam size centered at the CS(2-1) peak emission.
The measured values for T$_{MB}$(CS) and T$_{MB}$(C$^{34}$S) are listed in Table \ref{tbl:coldensity}.  
We use a terrestrial value of 22.5 for the CS to C$^{34}$S ratio \citep[e.g.,][]{1986PASJ...38..793K}. 
We further estimate the excitation temperature from the radiative transfer equation 
$$
T_{ex} = \frac{h\nu}{k} \left[ ln \left( \frac{h\nu / k}{\frac{T_{MB}}{1-e^{-\tau}}+ J_{\nu}(T_{bg})} +1 \right) \right] ^{-1},
$$ 
where $J_{v}(T) = \dfrac{h\nu / k}{e^{h\nu / kT}-1}$, $\nu$ is the frequency of the molecular transition, and T$_{bg}$ is the background temperature assumed to be 2.73 K.
With the assumption of LTE, the column density can be derived 
\[
\begin{aligned}
\left[ \frac{N}{\rm cm^{-2}} \right] =  & 1.67\times10^{14}\frac{Q_{rot}}{g_{k}g_{I}} \left[ \frac{S_{JKI}}{\rm erg \ cm^{3} \ statC^{-2} \ cm^{-2}} \right]^{-1} \\
     & \times \left[\frac{\mu}{\rm D} \right]^{-2} e^{Eu/T_{ex}} \left[ \frac{\nu}{\rm GHz} \right]^{-1} \frac{J_{v}(T_{ex})}{J_{v}(T_{ex})-J_{v}(T_{bg})} \\
     & \times \frac{\tau}{1-e^{-\tau}} \left[ \frac{\int T_{MB}dv}{\rm K \ km s^{-1}} \right].
\end{aligned}
\]
The parameters used in the equation are listed in Table \ref{tbl:csprop} \citep{2000tra..book.....R}.
We assume an abundance ratio between CS(2-1) and H$_{2}$ at the center of a core to be $2\times10^{-10}$ (the result from modeling in a later section) in converting the CS column densities to H$_{2}$ column densities.
Using a constant abundance ratio instead of a profile with central depletion is justified since the cores are chemically less evolved (see Sect.\ \ref{sect:iram}).
\citet{frau2010} reported similar CS-to-H$_{2}$ abundance ratio (few times $10^{-10}$) in the chemically young starless cores in the pipe nebula.

We further estimate the masses and number densities within the beam (25\arcsec) for the core peak assuming sphericity:
$$
M_{H_{2}} = \mu m_{H} D^{2} \int N_{H_{2}} d\Omega, 
$$
$$
n_{H_{2}} = \frac{M_{H_{2}}}{4\pi r^{3} /3},
$$
where $m_{H}$ is the hydrogen mass, D is the distance to Orion (414 pc from \citet{2007A&A...474..515M}), and r is the beam radius.  
The derived values for the optical depths, excitation temperatures,
column densities, and LTE masses at the peak positions are listed in
Table \ref{tbl:coldensity}. 

The optical depths of CS(2-1) are between 2 to 5; similar optical depths for CS(2-1) are reported in other star-forming cores \citep{frau2010,morata2003}.  
The excitation temperatures range from 6.5 K to 19 K with the average temperature of 11.3 K, in agreement with the kinetic temperature of 10 K (to 15 K) for starless cores \citep[e.g.,][]{2009ApJ...691.1754S}.
This suggests that the cores are thermalized; the thermalization is also suggested by the high number density (several times $10^{6}$ cm$^{-3}$) compared to the critical density for CS ($\sim 10^{5}$ cm$^{-3}$; see \citet{1999ARA&A..37..311E}).
The LTE masses indicate that these cores are massive star-forming regions. 
However, the LTE masses are inconsistent with the dust mass estimates from the 850 $\um$ observations, 
possibly due to the higher temperature (20 K) used in the dust masses calculation, the assumed dust opacity, the imperfect coupling between dust and gas, and optical depth effects.   
Also, due to the large-scale emission surrounding the cores, the masses here may be overestimated.  

The derived column densities ($\ge 10^{23}$ cm$^{-2}$) are comparable to several massive star-forming regions \citep[e.g.,][]{2007A&A...466.1065B} and are slightly higher than some intermediate mass-star forming regions \citep[e.g.,][]{2011MNRAS.415.2790L}.  
The number densities range from $4.7\times 10^{6}$ to $3.7\times 10^{7}$ cm$^{-3}$.  
The simulation performed by \citet{2005ApJ...635.1151K} which includes hydrodynamics, radiative cooling, variable molecular abundance, and radiative transfer concludes that starless cores with central densities larger than a few times $10^{5}$ cm$^{-3}$ are dynamically unstable and may proceed to gravitational collapse, 
suggesting that our cores are gravitationally bound.

\begin{figure*}
\begin{center}
\ContinuedFloat
\includegraphics[scale=0.7,angle=0]{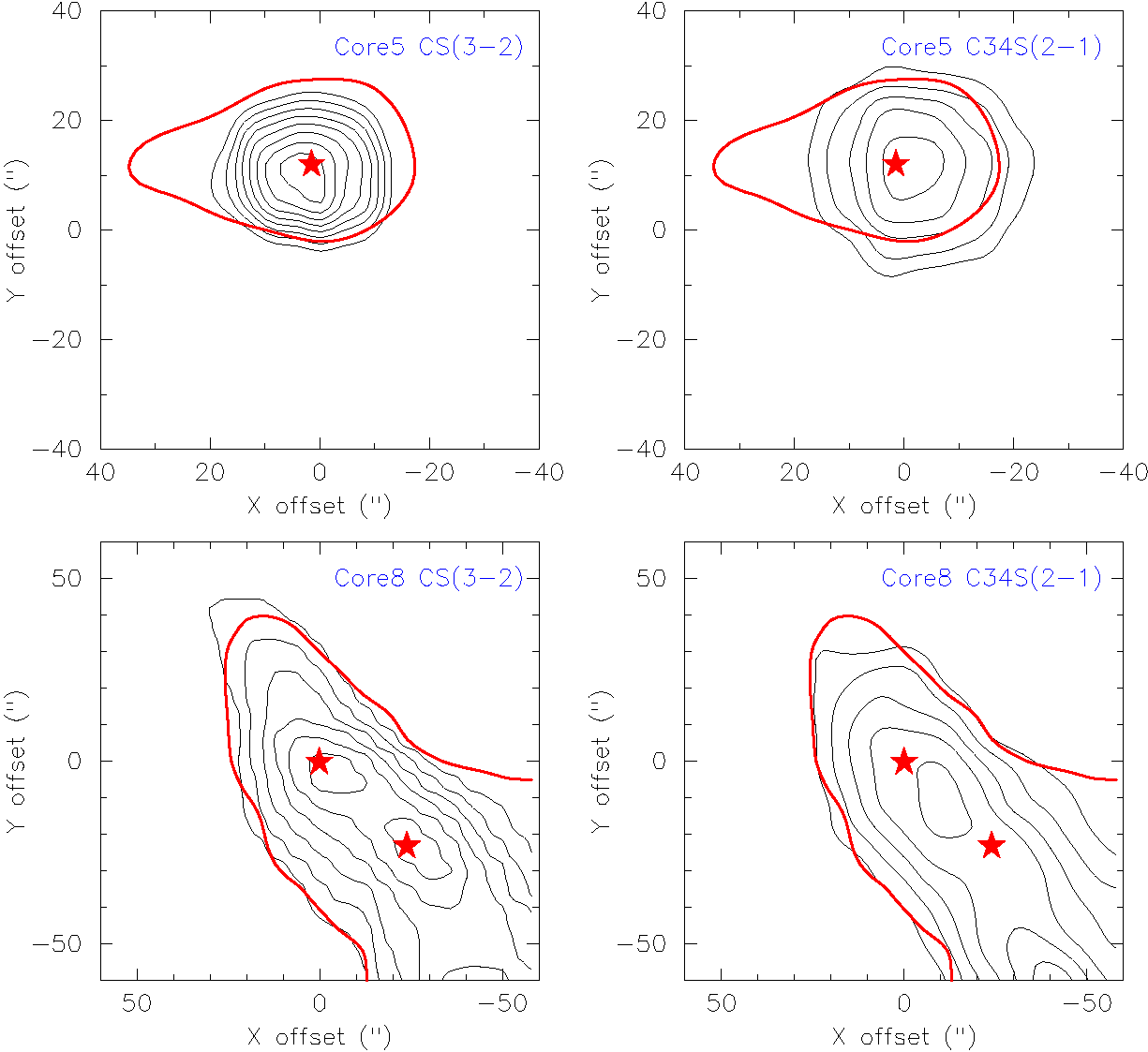}
\caption{continued.}
\end{center}
\end{figure*}

\subsection{CARMA maps}
\subsubsection{CARMA CS(2-1) maps}
\label{sect:carma_cs21} 

\begin{deluxetable}{llc}
\tablecaption{CS(2-1) Properties}
\tabletypesize{\footnotesize}
\tablewidth{240pt}
\tablecolumns{3}
\tablehead{\colhead{Parameter} & \colhead{Description} & \colhead{Value}}
\startdata
Q$_{rot}$ & rotational partition function & 0.86T$_{ex}$ \\
g$_{I}$, g$_{k}$ & degeneration of the quantum number & g$_{I}$=1, g$_{k}$=1 \\
S$\mu^{2}$      & S: line strength, $\mu^{2}$: dipole moment & 7.71 debye       \\
E$_{u}$ &       energy in the upper state       & 7.0 K \\
$\nu$   &       frequency of the molecular line & 97.980968 GHz
\enddata
\label{tbl:csprop}
\end{deluxetable}

Of the 9 cores detected by the IRAM 30-m, 8 of them were followed up by higher resolution CARMA observations (Figure \ref{fig:fragment}).
Most of the cores (except Core 5) show multiple intensity peaks in the CARMA CS(2-1) emission within the single peaked IRAM cores, suggesting fragmentation.
These fragments are not in spherical morphologies and are spatially connected to each other.  
The number of fragments range from three to five in each core.  
Core 5 is associated with one single object in spherical shape and does not show signs of fragmentation.  
 
The CARMA CS(2-1) emission well traces the small-scale filamentary structures probed by the JCMT SCUBA-2 850 $\um$ observations and Herschel 500 $\um$ observations.
For example, Core 1 is associated with the filamentary structure in the North-East and South-West direction as clearly seen in the 850 $\um$ and 500 $\um$ images.
The tight connection between the structures probed by CS(2-1) and dust emission is reminiscent of the star formation activities along filamentary structures at large scales \citep[see][]{lee2012}, 
suggesting the importance of filamentary structures to star formation at small scales.

\begin{figure*}
\begin{center}
\includegraphics[scale=0.65,angle=270]{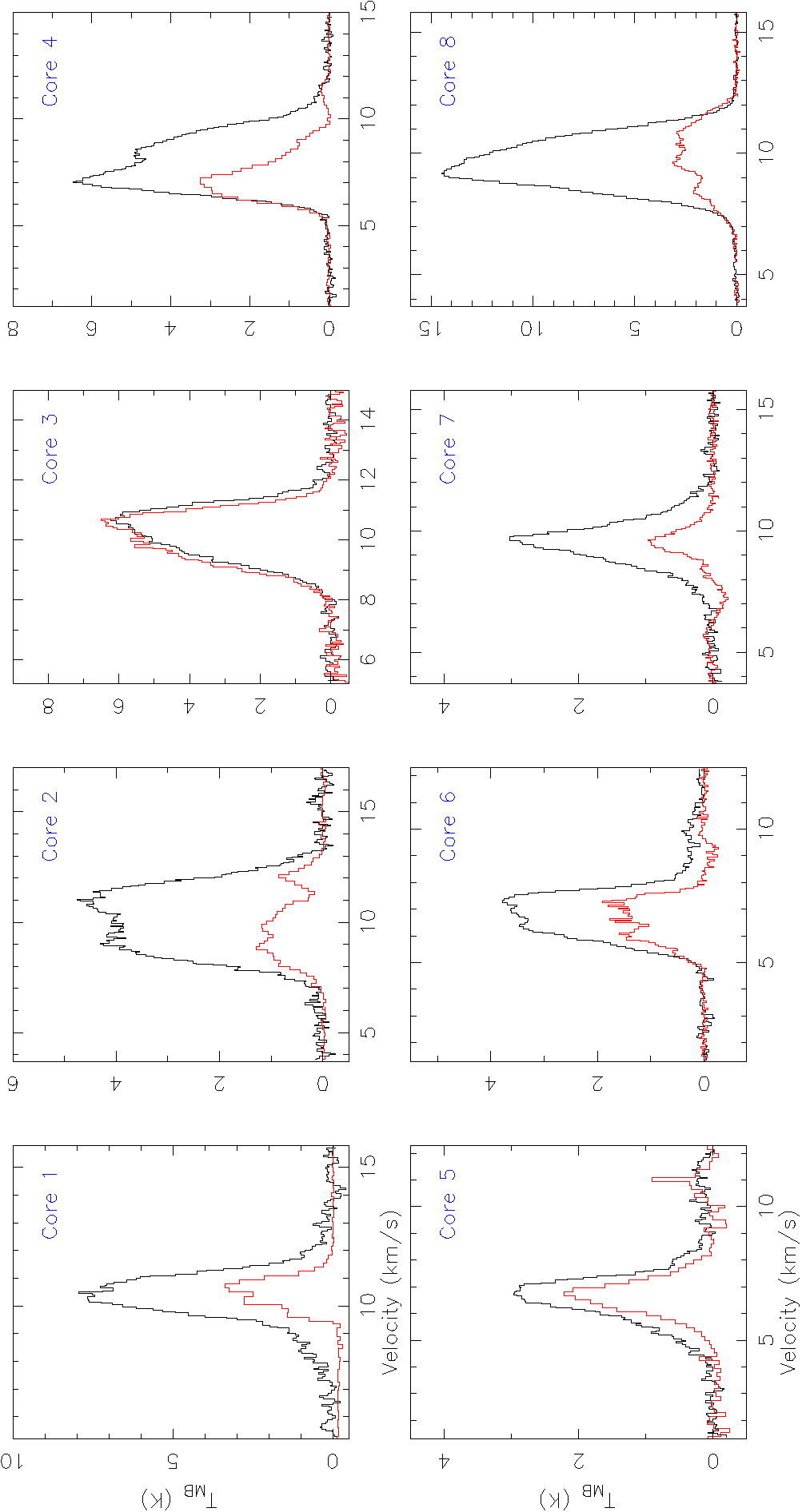}
\caption{Comparison between the IRAM CS(2-1) flux density (black lines) and the CARMA CS(2-1) flux (red lines).
The RMS noises for the IRAM data are: 0.24 K (Core 1), 0.15 K (Core 2), 0.1 K (Core 3), 0.085 K (Core 4), 0.1 K (Core 5), 
0.09 K (Core 6), 0.075 K (Core 7), and 0.075 K (Core 8).  
The RMS noises for the convolved CARMA data are: 
0.096 K (Core 1), 0.17 K (Core 2), 0.17 (Core 3), 0.094 K (Core 4), 0.087 K (Core 5), 
0.093 (Core 6), 0.116 K (Core 7), 0.122 K (Core 8). 
}
\label{fig:resolve}
\end{center}
\end{figure*}

We compare the CS(2-1) fluxes from the IRAM observations and the CARMA observations for all 8 cores (Figure \ref{fig:resolve}).
For this comparison, the CARMA maps are convolved with the beam size of the IRAM 30-m telescope (25\arcsec).
For both IRAM and CARMA fluxes, the spectra are then extracted from the averaged flux within the lowest contour level for each core shown in Figure \ref{fig:cs}.
Some of the cores (Core 1, 2, 4, 7 and 8) show that the CARMA fluxes are largely resolved out while the other cores (Core 3, 5 and 6) show comparable CARMA and IRAM fluxes.
The largely resolved out fluxes with Core 1, 2, 4, and 8 may be associated with converging flows at larger scales (see Sect.\ \ref{sect:radiative}).

\begin{figure*}
\begin{center}
\includegraphics[scale=0.7,angle=0]{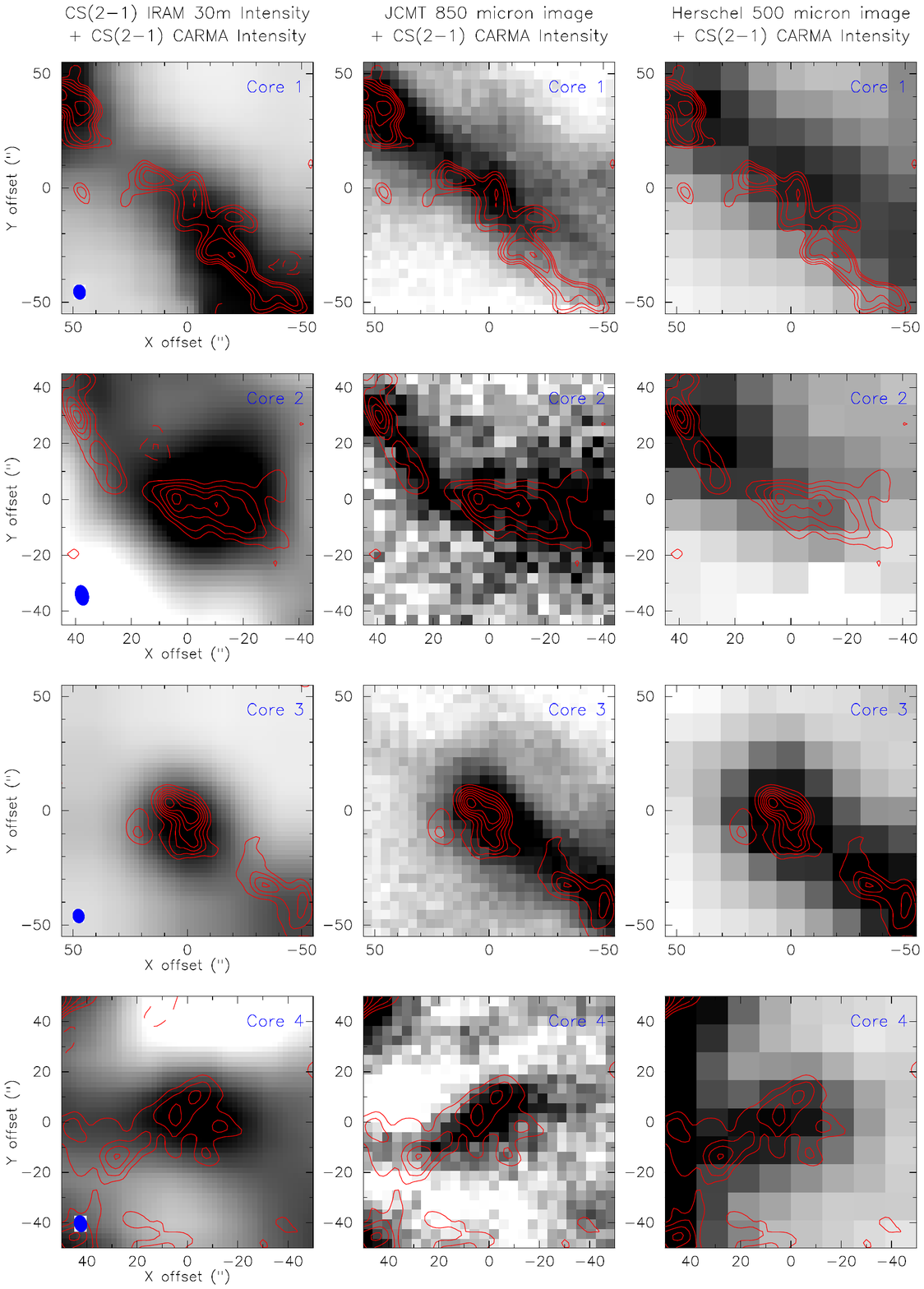}
\caption{The $0^{th}$ moment maps from the CARMA CS(2-1) observations (red contours) overlaid on the IRAM CS(2-1) observations (first column),
JCMT 850 $\um$ dust continuum (second column), and the \textit{Herschel} 500 $\um$ dust continuum (third column).
The contour levels ($\sigma$, starting level, interval) are:
Core 1: $\sigma=0.32$ Jy beam$^{-1}$ km s$^{-1}$, $\pm$5$\sigma$, $\times \sqrt{2} \sigma$;
Core 2: $\sigma=0.32$ Jy beam$^{-1}$ km s$^{-1}$, $\pm$5$\sigma$, $\pm$2$\sigma$;
Core 3: $\sigma=0.45$ Jy beam$^{-1}$ km s$^{-1}$, $\pm$5$\sigma$, $\pm$5$\sigma$;
Core 4: $\sigma=0.4$ Jy beam$^{-1}$ km s$^{-1}$, $\pm$5$\sigma$, $\pm$3$\sigma$;
Core 5: $\sigma=0.18$ Jy beam$^{-1}$ km s$^{-1}$, $\pm$5$\sigma$, $\pm$5$\sigma$;
Core 6: $\sigma=0.16$ Jy beam$^{-1}$ km s$^{-1}$, $\pm$10$\sigma$, $\pm$10$\sigma$;
Core 7: $\sigma=0.2$ Jy beam$^{-1}$ km s$^{-1}$, $\pm$5$\sigma$, $\pm$2$\sigma$;
Core 8: $\sigma=0.6$ Jy beam$^{-1}$ km s$^{-1}$, $\pm$10$\sigma$, $\pm$3$\sigma$.
}
\label{fig:fragment}
\end{center}
\end{figure*}

\begin{figure*}
\begin{center}
\ContinuedFloat
\includegraphics[scale=0.7,angle=0]{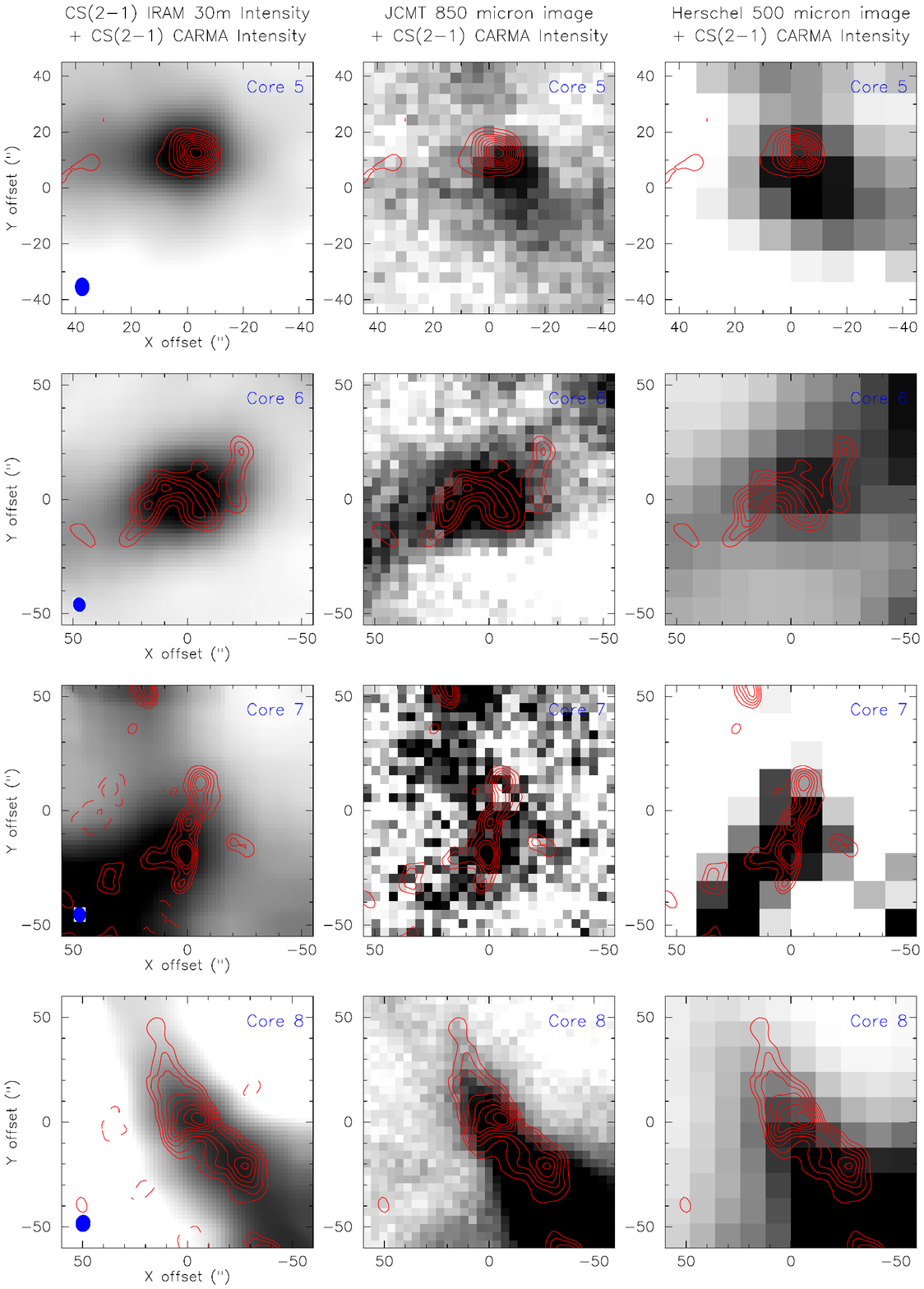}
\caption{Continued.}
\end{center}
\end{figure*}

\subsubsection{CARMA N$_{2}$H$^{+}$(1-0) maps}

\begin{figure*}
\begin{center}
\includegraphics[scale=0.7,angle=0]{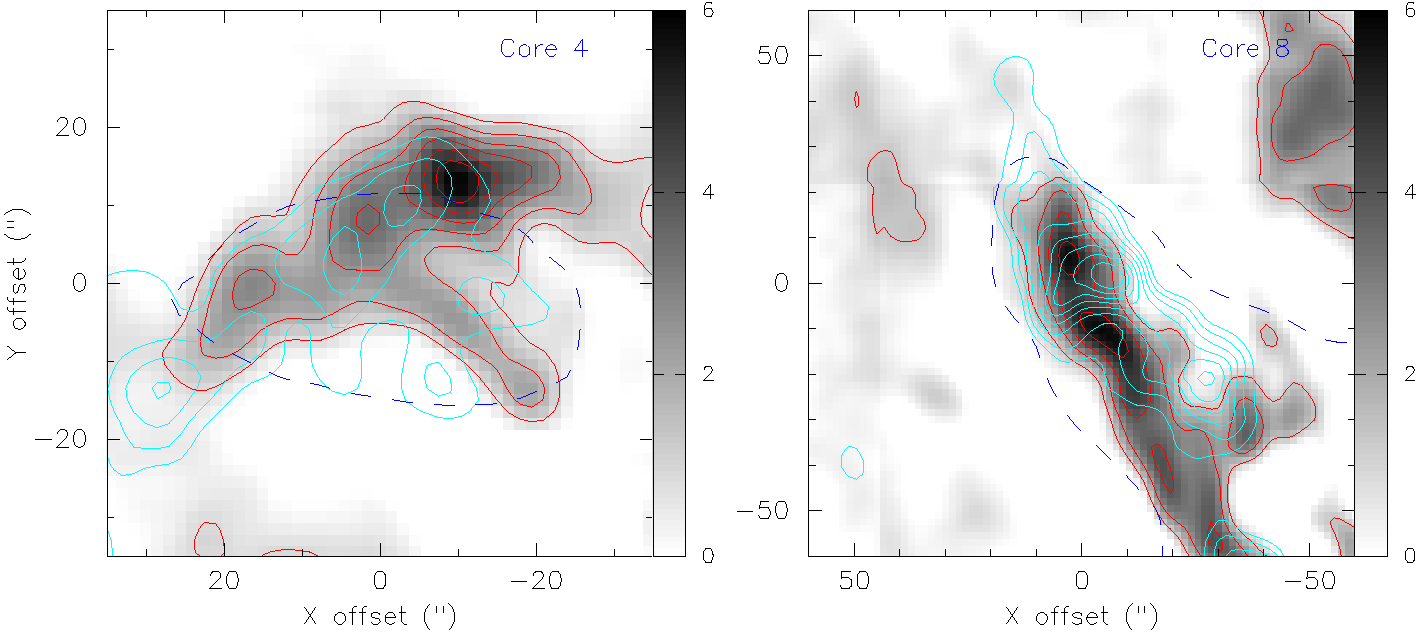}
\caption{The $0^{th}$ moment maps from the CARMA N$_{2}$H$^{+}$(1-0) observation (red contours and grey scale) in comparison with the $0^{th}$ moment maps from the CARMA CS(2-1) observations (cyan contours).  The blue dashed lines are the IRAM CS contour.
The contour levels for the N$_{2}$H$^{+}$(1-0) emission are 10\%, 20\%, 30\%, 40\%, 50\%, 60\%, 70\%, 80\%, 90\%, and 100\% of the peak intensities (in Jy beam$^{-1}$ km s$^{-1}$).
}
\label{fig:n2h+}
\end{center}
\end{figure*}

N$_{2}$H$^{+}$(1-0) is detected only in Core 4 and Core 8 with CARMA while the other cores show no detections (Figure \ref{fig:n2h+}).
For Core 4, N$_{2}$H$^{+}$(1-0) shows multiple peaks with the strongest emission outside the IRAM contour.
For Core 8, N$_{2}$H$^{+}$(1-0) also shows two major peaks.
Although the positions of N$_{2}$H$^{+}$(1-0) peaks do not well coincide with the CS(2-1) peaks for both cores,
N$_{2}$H$^{+}$(1-0) still shows the clumpy nature for these cores, suggesting that the fragmentation detected by optically thicker tracer CS(2-1) is not due to the chemical effect from depletion.

\subsubsection{CARMA continuum maps at 3 mm}
There was no detections in the CARMA continuum at 3mm.
The non-detection is reminiscent of the lack of 3 mm continuum emission in CARMA D array maps towards 9 starless cores in the Perseus molecular cloud \citet{2010ApJ...718..306S}.
\citet{lee2012} suggested that the non-detections in the Perseus sample are probably due to a combination of resolving-out structure and sensitivity.  
Future observations with better sensitivity are required for further characterization on dust properties at millimeter-wavelengths.
We calculated the mass sensitivity from the 3$\sigma$ upper limit using the equation:
\begin{equation}
M = \frac{d^{2}S_{3mm}}{B_{\nu}(T_{D})\kappa_{3mm}}
\end{equation}
where d is distance to Orion, $S_{3mm}$ is three times the noise level, $B_{\nu}$ is the Planck function as a function of dust temperature $T_{D}$, and $\kappa_{3mm}$ is the dust opacity. 
We use a typical dust temperature of 10 K for starless cores and 0.00169 cm$^{2}$ g$^{-1}$ for $\kappa_{3mm}$ (an extrapolated value from \citep{1994A&A...291..943O}) by assuming a gas-to-dust ratio of 100 and $\beta=2$.  
The noise levels for the cores and the mass upper limits corresponding to $3\sigma$ are presented in Table \ref{tbl:sensitivity}.
These mass limits are only for substructures and the large-scale emission is not detected.

\begin{deluxetable}{lcc}
\tablecaption{Noise Level for 3 mm continuum observations and Mass Sensitivity}
\tabletypesize{8pt}
\tablewidth{240pt}
\tablecolumns{3}
\tablehead{
\colhead{Source} & \colhead{Noise} & \colhead{Mass\tablenotemark{a}} \\
\colhead{} & \colhead{(mJy beam$^{-1}$)} & \colhead{(M$_{\sun}$)} 
}
\startdata
Core 1 & 4.7 & $<$ 2.85 \\
Core 2 & 1.3 & $<$ 0.79 \\
Core 3 & 7.7 & $<$ 4.67 \\
Core 4 & 1.9 & $<$ 1.15 \\
Core 5 & 2.2 & $<$ 1.33 \\
Core 6 & 2.4 & $<$ 1.42 \\
Core 7 & 15.0 & $<$ 9.09 \\
Core 8 & 1.9 & $<$ 1.15 
\enddata
\label{tbl:sensitivity}
\tablenotetext{a}{The mass corresponds to the 3$\sigma$ upper limit for detection of small-scale structure.}
\end{deluxetable}

\section{Results and Data Analysis II: Kinematics}

\subsection{Large-scale Kinematics with IRAM}

\subsubsection{Velocity Gradients Fitting}

\begin{figure*}
\begin{center}
\includegraphics[scale=0.85,angle=0]{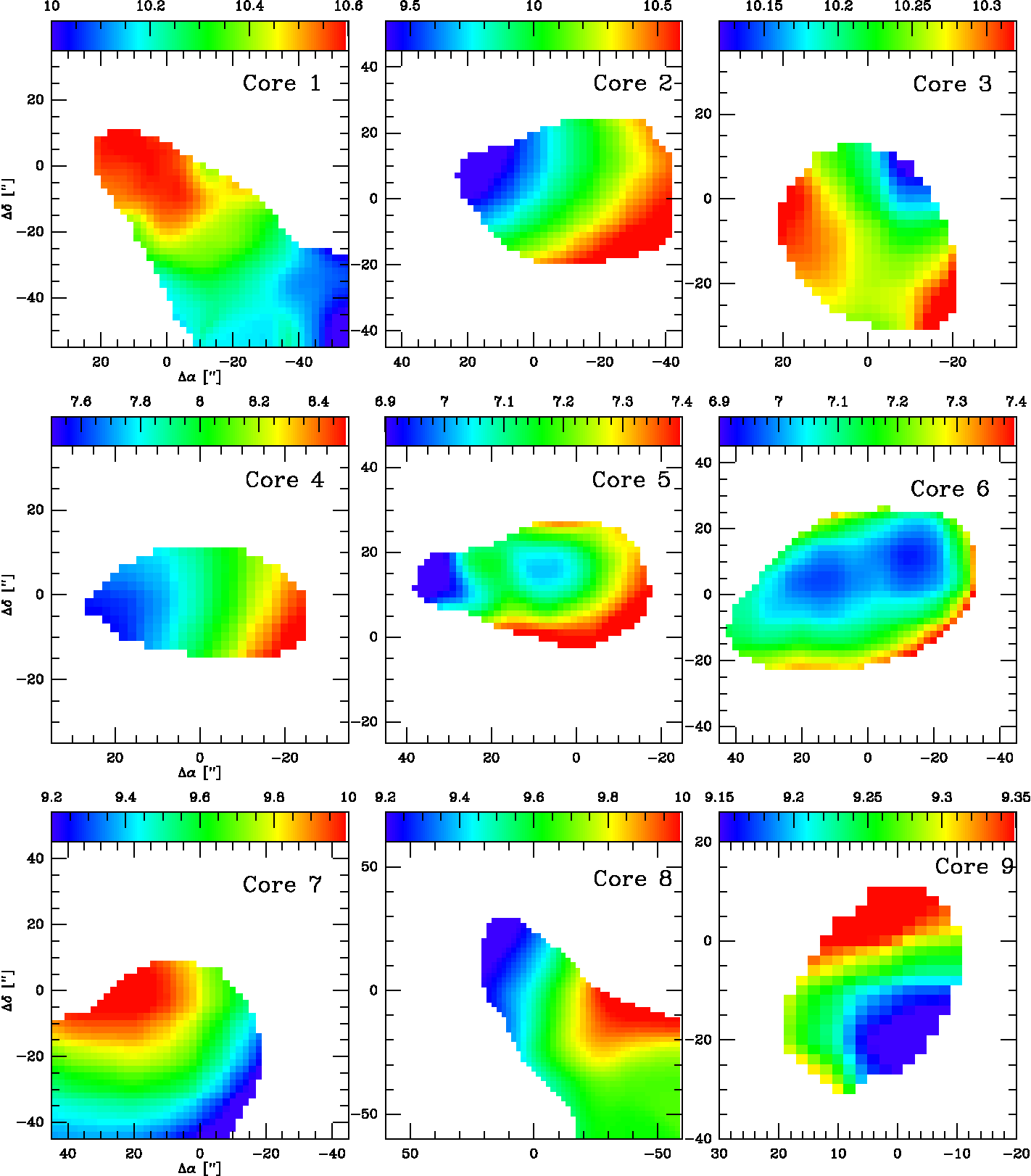}
\caption{First moment maps from the IRAM CS(2-1) data.  The color scale indicates velocity in km s$^{-1}$.}
\label{fig:iram_mom1}
\end{center}
\end{figure*}

Figure \ref{fig:iram_mom1} shows the first moment maps from the IRAM CS(2-1) data for the nine cores, masked
based on the CS(2-1) contours. 
Velocity gradients are observed in several cores (Core 1, 2, 4, 7, 8, and 9).
To derive the magnitudes of the velocity gradients, we assume that the gradients are linear in both right ascension and declination.
Using the first-moment maps, the velocity gradients are computed based on the method described in \citet{1993ApJ...406..528G}, 
but with the MPFIT function \citep{2009ASPC..411..251M} implemented in
IDL performing the least-square fitting:  
\begin{equation}
v_{LSR} = v_{0} + a \Delta \alpha + b \Delta \delta, 
\end{equation}
where $v_{0}$ is the systematic velocity, $a$ and $b$ are the velocity gradients along the right ascension and declination, 
and $\Delta \alpha$ and $\Delta \delta$ are the position offsets.  
The total velocity gradients are defined as 
$$
g = \frac{\sqrt{a^{2}+b^{2}}}{D},
$$
where D is the distance (414 pc for Orion).
The direction of the velocity gradients is then defined as $\theta_{g}=tan^{-1}(\frac{b}{a})$.

The fitting results of the velocity gradients for all nine cores are listed in Table \ref{tbl:gradients}. 
The total velocity gradients range from 1.4 to 12.1 km s$^{-1}$ pc$^{-1}$, with an average value of 6.2 km s$^{-1}$ pc$^{-1}$.
Most of the velocity gradients are large compared to dense cores including starless cores and protostars 
(0.3 to 4 km s$^{-1}$ pc$^{-1}$ from \citet{1993ApJ...406..528G} and 0.5 to 6 km s$^{-1}$ pc$^{-1}$ in \citet{2002ApJ...572..238C}).
Recent observations with higher resolutions have found larger velocity gradients.  
For example, \citet{2011MNRAS.410...75C} found the velocity gradients of $\sim$ 5.5 km s$^{-1}$ pc$^{-1}$ for starless cores and $\sim$ 6 km s$^{-1}$ pc$^{-1}$ for protostars in Perseus.
Also, \citet{2011ApJ...740...45T} found a median velocity gradient of 10.7 km s$^{-1}$ pc$^{-1}$ with several Class 0 objects from interferometric data.

These velocity gradients are often interpreted as rotation.
However, the common interpretation of rotation needs to be treated with caution 
since an inflowing filament can also produce the velocity patterns that mimic rotation \citep{2012ApJ...748...16T}.
Infall and rotation in spherical objects are easy to distinguish
since spherical infall exhibits blue-skewed spectra with optically thick lines across the object \citep[e.g.,][]{1999ApJ...526..788L,2008ApJ...689..335P}. 
However, infall and rotation in filaments are more difficult to disentangle as both generate velocity gradients.
The spectral maps and position-velocity diagrams need to be examined carefully to correctly interpret the kinematics.

\begin{deluxetable*}{clccccc}
\tablecaption{2D fitting of velocity gradients}
\tabletypesize{\footnotesize}
\tablewidth{0pt}
\tablecolumns{6}
\tablehead{
\colhead{Source} & \colhead{v$_{0}$} & \colhead{a} & \colhead{b} & \colhead{g} & \colhead{$\theta_{g}$}  \\
\colhead{} & \colhead{(km s$^{-1}$)} & \colhead{(km s$^{-1}$ pc$^{-1}$)} & \colhead{(km s$^{-1}$ pc$^{-1}$)} & \colhead{(km s$^{-1}$ pc$^{-1}$)} & \colhead{(degree)}
}
\startdata
Core 1 & 10.49$\pm$0.20  & 3.43$\pm$0.35  & 1.17$\pm$0.43  & 4.28$\pm$0.42 & 71.9$\pm$5.7  \\
Core 2 & 9.91$\pm$0.16  & -9.96$\pm$0.51  & -6.91$\pm$0.76 & 14.31$\pm$0.71 & 55.2$\pm$2.8  \\
Core 3 & 10.25$\pm$0.20  & 0.87$\pm$0.45  & -1.11$\pm$0.41 & 1.66$\pm$0.51 & 321.7$\pm$17.3  \\
Core 4 & 7.98$\pm$0.19  & -8.87$\pm$0.52  & -1.76$\pm$0.92 & 10.67$\pm$0.64 & 78.79$\pm$3.4  \\
Core 5 & 7.28$\pm$0.19  & -4.31$\pm$0.84  & -8.07$\pm$1.40 & 10.81$\pm$1.52 & 28.10$\pm$8.1  \\
Core 6 & 7.07$\pm$0.12  & -0.43$\pm$0.45  & -1.36$\pm$0.58 & 1.69$\pm$0.61 & 17.42$\pm$22.9  \\
Core 7 & 9.77$\pm$0.20  & 2.67$\pm$0.46   & 6.73$\pm$0.48  & 8.54$\pm$0.55 & 21.63$\pm$3.8 \\
Core 8 & 9.45$\pm$0.10  & -6.01$\pm$0.34  & 0.18$\pm$0.31  & 7.09$\pm$0.40 & 271.8$\pm$3.2 \\
Core 9 & 9.29$\pm$0.28  & 3.68$\pm$0.80   & 3.77$\pm$0.58  & 6.22$\pm$0.83 & 44.3$\pm$7.6
\enddata
\label{tbl:gradients}
\end{deluxetable*}

\begin{figure*}
\begin{center}
\includegraphics[scale=0.7,angle=270]{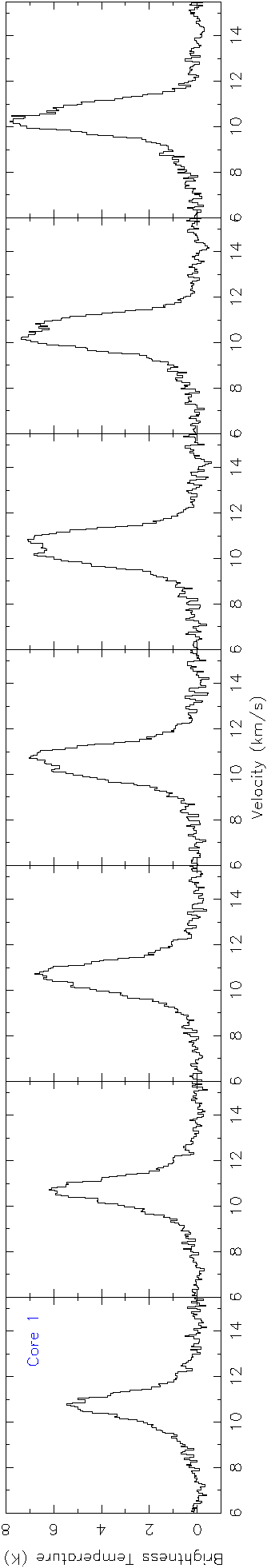} \par
\vspace{0.18cm}
\includegraphics[scale=0.7,angle=270]{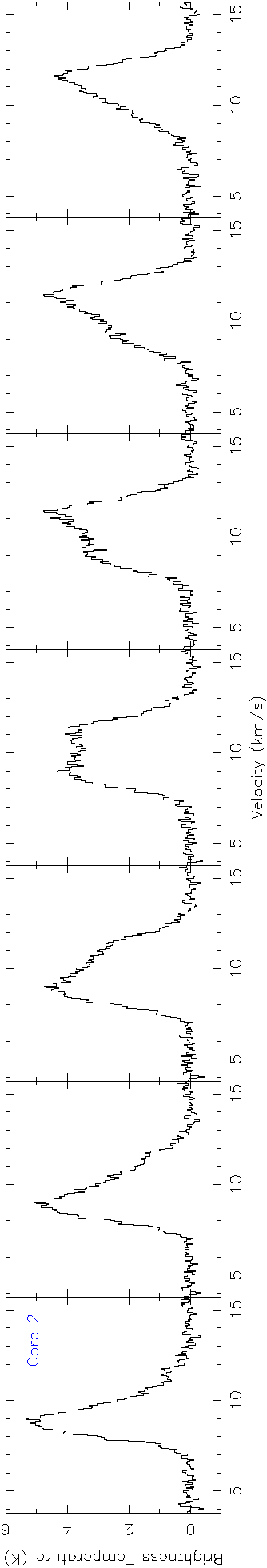} \par
\vspace{0.18cm}
\includegraphics[scale=0.65,angle=270]{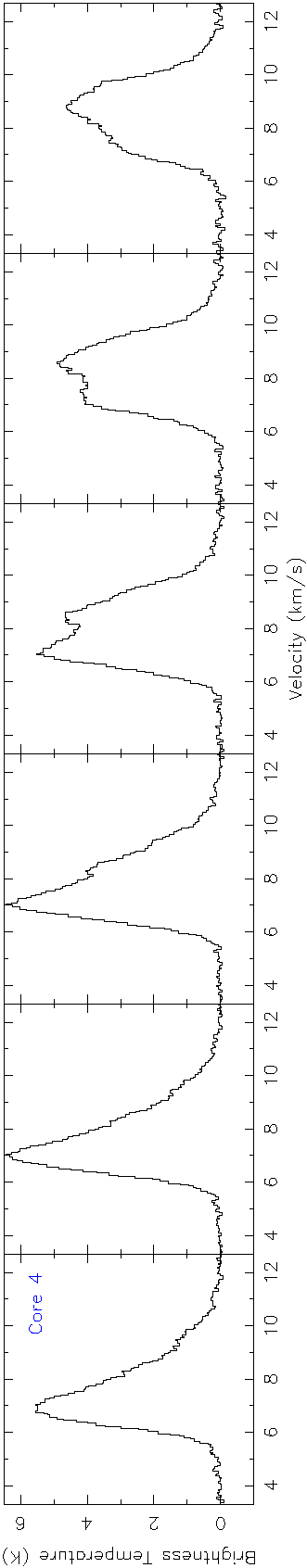} \par
\vspace{0.18cm}
\includegraphics[scale=0.65,angle=270]{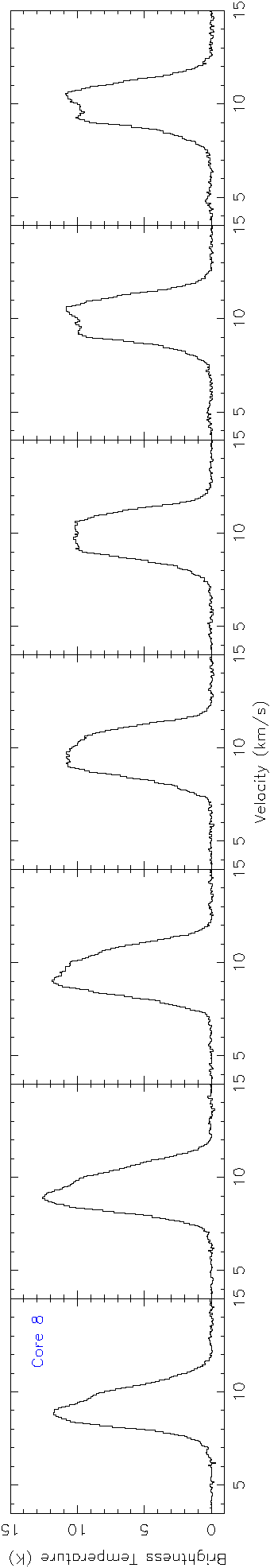}
\caption{Spectra of the four cores along the directions of the velocity gradients.}
\label{fig:spec}
\end{center}
\end{figure*}

\subsubsection{Spectral Maps}
\label{sect:spectra_cs21}

Cores 1, 2, 4, and 8 have the most prominent velocity gradients in the IRAM 30-m data.
By using a linear cut across the minimum and maximum velocities in the maps, one notices two main features of the spectra across these cores
(Figure \ref{fig:spec}).
First, a gradual shift from blue to red in the peak velocities is observed across the cores.
The separations between the blue-shifted and red-shifted peak velocities are $\sim$ 0.8 km s$^{-1}$ (Core 1), 2.5 km s$^{-1}$ (Core 2), 1.6 km s$^{-1}$ (Core 4), and 1.7 km s$^{-1}$ (Core 8).   
Also, 
The velocity peaks of the blue components and that of the red components do not vary significantly across the cores, suggesting that the gas flows at a nearly constant speed.
Second, the blue components have stronger intensities than the red components.
These two features are observed for all four cores.
To explain these two features, we performed a radiative transfer modeling of Core 2.

\subsubsection{Radiative Transfer Modeling}
\label{sect:radiative}

\begin{figure*}
\begin{center}
\includegraphics[scale=0.6]{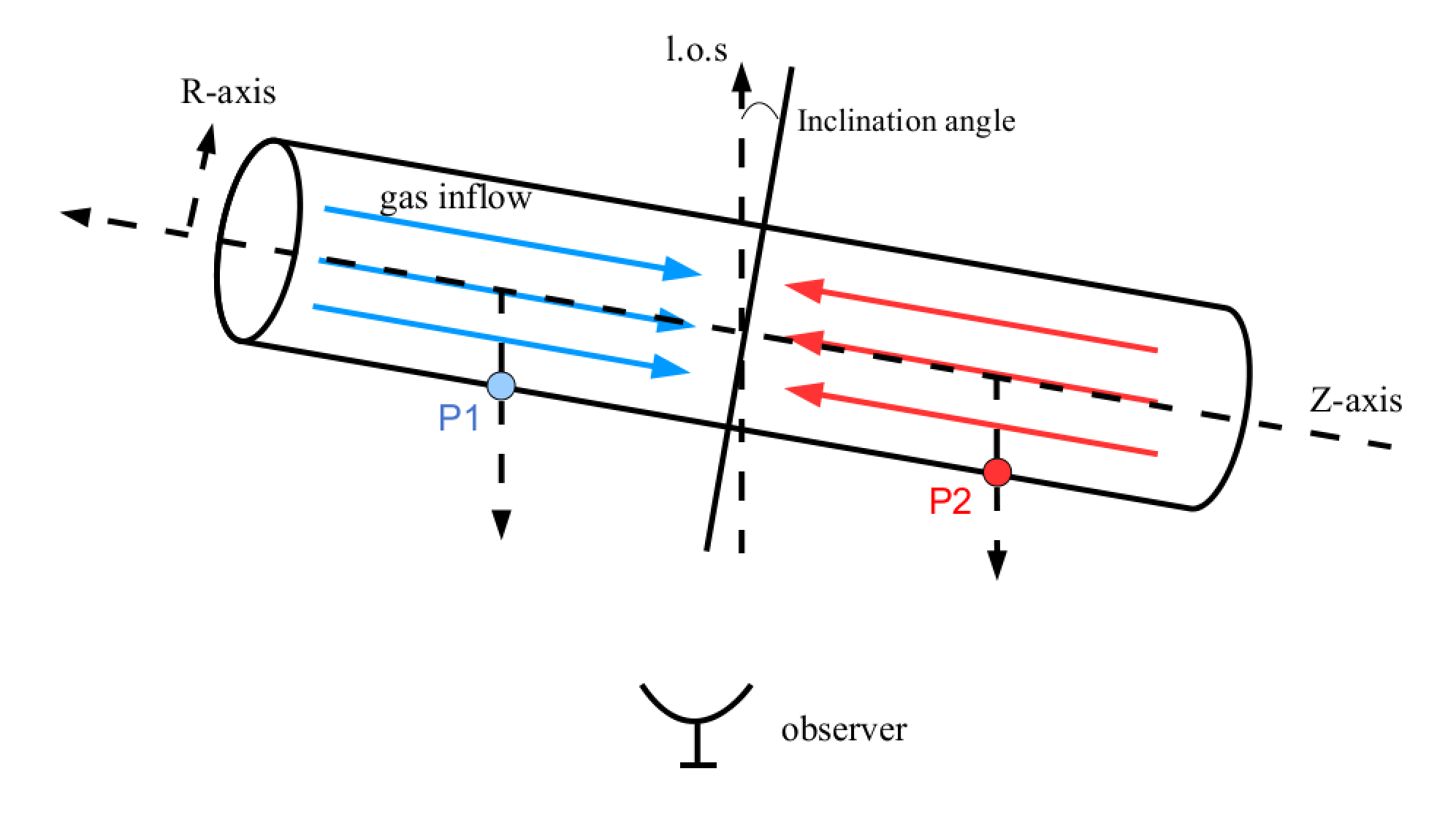}
\caption{The model for the radiative transfer modeling.}
\label{fig:model}
\end{center}
\end{figure*}

\begin{figure*}
\begin{center}
\includegraphics[scale=0.7,angle=270]{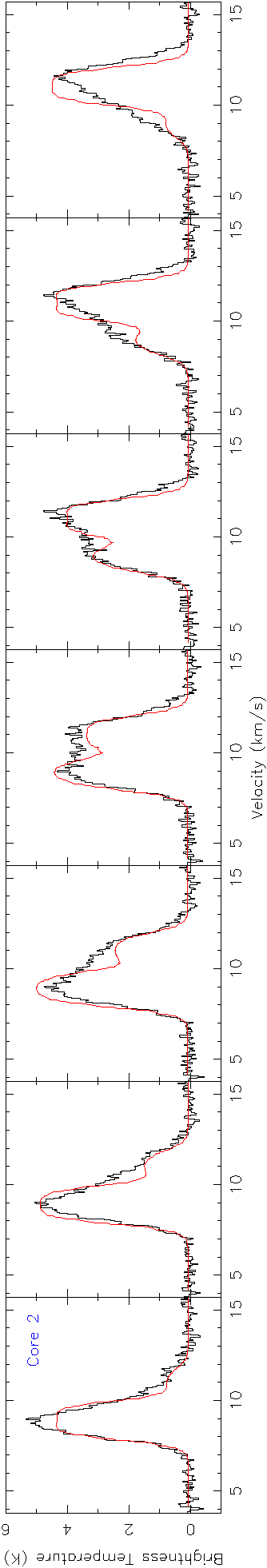} \par
\vspace{0.18cm}
\includegraphics[scale=0.7,angle=270]{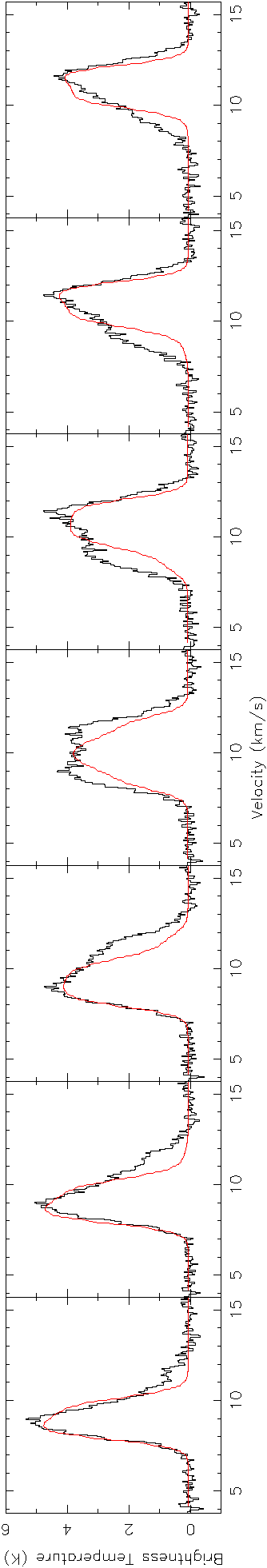}
\caption{The inflow model (top) and the rotation model (bottom).  The black lines are from the data of Core 2 and the red lines indicate the radiative transfer model.}
\label{fig:rotation}
\end{center}
\end{figure*}

We compare the CS(2-1) spectra from Core 2 with radiative transfer calculations performed with the code \textit{LIME} (Line Modeling Engine; \citet{2010A&A...523A..25B}). 
We chose Core 2 to model since it best presents the two common features observed for the four cores.  
The code calculates the emergent spectra by solving the molecular line excitation with 3D Delaunay grids for photon transport and accelerated Lambda Iteration for population calculations.
The inputs to the code, which are based on 3D structures, are the density, temperature, chemical abundance, velocity, and linewidth profiles.  
Users also control parameters such as molecular lines, inclination angles, dust properties, and image resolutions. 

To be consistent with the flattened morphologies of the cores, we consider a cylindrically symmetric filament that contains inflowing\footnotemark[7] gas from the two sides to the center with an inclination angle (Figure \ref{fig:model}), which is similar to filament models \citep[e.g.,][]{peretto2006}.
The density is considered to vary with the cylindrical R and Z.
The density profile has the form of $n(R,Z) = n_{0} / [1 + (\frac{R}{R_{0}})^{\alpha}] / [1 + (\frac{Z}{Z_{0}})^{\alpha}]$,
where $n_{0}$, $R_{0}$ and $Z_{0}$ are constant.
The power-law index $\alpha$ is taken to be 2.5 in the modeling, consistent with other starless cores \citep{2002ApJ...569..815T,2004A&A...416..191T}.
We varied $n_{0}$ with $3.0 \times 10^{6}$, $4.0 \times 10^{6}$, and $5.0 \times 10^{6}$ cm$^{-3}$;
$R_{0}$ and $Z_{0}$ were varied with four sets of numbers respectively: (4180 AU, 17333 AU), (5513 AU, 30000 AU), (6180 AU, 35066 AU), and (10000 AU, 43333 AU).

\begin{figure*}
\begin{center}
\includegraphics[scale=0.7,angle=270]{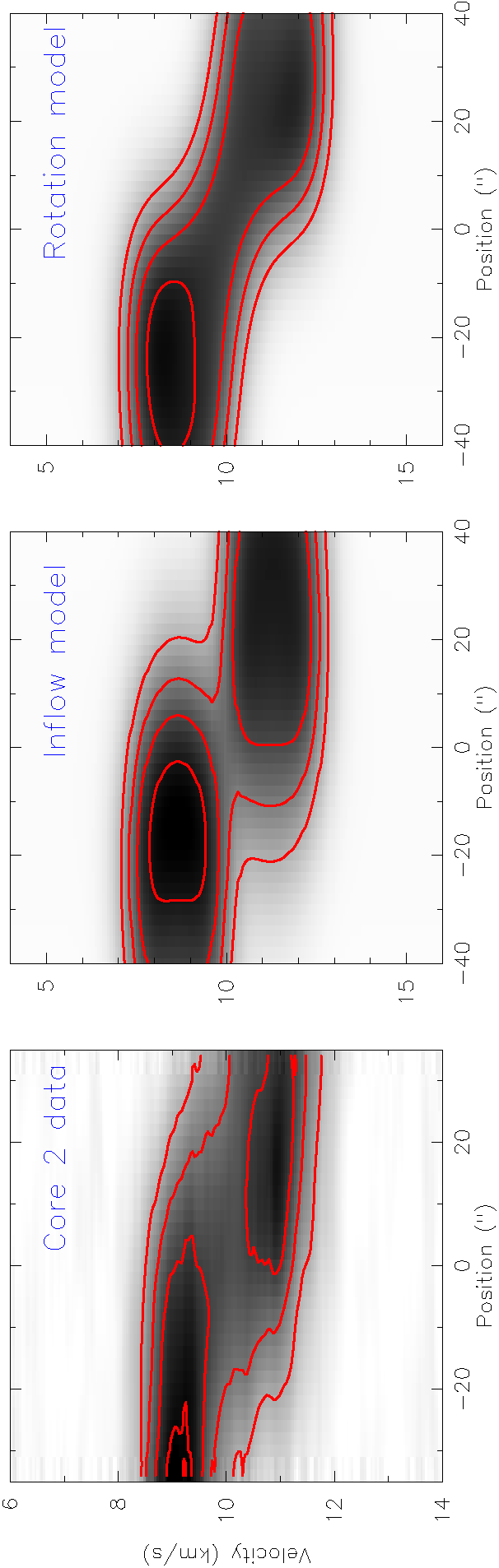}
\caption{The PV diagrams of the data (left panel), inflow model (central panel), and rotation model (right panel).  The PV diagrams are from the cut shown in Fig.\ \ref{fig:spec}.  The contours are 30\%, 50\%, 70\%, and 90\% of the peak values in all three panels.}
\label{fig:pv}
\end{center}
\end{figure*}

The temperature is assumed to be a typical temperature of 10 K for starless cores \citep[e.g.,][]{2009ApJ...691.1754S}.
For the CS(2-1) abundance, we have considered constant abundance ratios ([CS]/[H$_{2}$] = $10^{-9}$, $3\times 10^{-10}$, $2\times 10^{-10}$) and a centrally depleted profile 
that has the abundance ratio of $10^{-11}$ in the center and $10^{-9}$ in the outer envelope. 
For the velocity field, we have considered two profiles in both the velocity fields that are along the Z-axis only.
First, we considered constant velocities including 1.3 km $s^{-1}$ and 1.5 km $s^{-1}$.
Second, we considered a profile that has the form of Keplerian rotation: $v(Z) = v_{0}(Z/Z_{c})^{-0.5}$, where $Z_{c}$ is a constant to modulate the profile.
$v_{0}$ was varied with 2.0 km $s^{-1}$, 3.0 km $s^{-1}$, and 4.0 km $s^{-1}$; $Z_{c}$ was varied with 500 AU and 2000 AU.
The line dispersion was considered with 0.5 km $s^{-1}$ and 0.8 km $s^{-1}$.
The inclination angles were tested with 18 degree, 45 degree, and 60 degree.

The best-fit gives a density profile of $n(R,Z) = 3.0 \times 10^{6} / [1 + (\frac{R}{10000 AU})^{2.5}] / [1 + (\frac{Z}{43333 AU})^{2.5}]$, 
a constant temperature of 10 K, a constant [CS]/[H$_{2}$] abundance ratio of $2.0 \times 10^{-10}$, a constant linewidth of 0.8 km $s^{-1}$, 
a constant inflow velocity of 1.5 km $s^{-1}$, and an inclination angle of 45 degree.
The total reduced $\chi^{2}$ is calculated to be 1.58.
The red line in the spectra of Core 2 in Figure \ref{fig:rotation} shows the best-fit from the radiative transfer modeling.
The top figure shows the result from the inflow model.  
The model has been convolved with the same beam size as the observational data.

The model successfully explains the two features we observed in the spectral maps.
With an inclination angle, the gas flow further from observers becomes blue-shifted and the side closer to observers becomes red-shifted.
Therefore, shifts of peak velocities from blue to red are observed.
For an optically thick line such as CS(2-1),
the emission produced in the blue-shifted side is closer to the symmetrical center (P1 in Figure \ref{fig:model}) than the emission produced in the red-shifted side due to the projection (P2 in Figure \ref{fig:model}).
The spectral intensity is higher closer to the symmetric center since the excitation temperature is higher due to the density profile.
As as result, we see the intensities in the blue-shifted side larger than the red-shifted side.
The reason for this asymmetry is similar to the blue-skewed spectra for spherical infall \citep[e.g.,][]{1999ARA&A..37..311E}.
However, we stress that the central dip seen in the model shown at the center of the core is not caused by self-absorption.
Instead, the dip is due to the overlapping between the two Gaussian velocity components. 
Self-absorption would occur at the inflow velocity for each of the velocity component;
however, although CS(2-1) is optically thick ($\tau$ $\sim$ 3) for our cores, we do not observe self-absorptions.

As mentioned above, 
it is considerably challenging to distinguish between 2D inflow or rotation on a filament from observations \citep{2012ApJ...748...16T}.
To examine the differences between the two compared with our data, 
we also performed the radiative transfer modeling on a cylindrically symmetric filament with rotation on the R- axis.  
All the profiles and parameters are the same as the best-fit model for filamentary inflow except for the velocity field.
The velocity field we adopted is a constant velocity field along the line-of-sight 
since the two peaks of the red and blue components nearly stay constant and the best-fit for the inflow model gives a constant velocity.
Therefore, the velocity field shows a profile of differential rotation where $\omega(z) \propto \frac{1}{z}$.

Figure \ref{fig:rotation} shows the spectra from the rotation model.  
The black lines are from the data of Core 2 and the red lines indicate the radiative transfer model.
As shown in the figure,
the model fails to describe not only the broad linewidth in the central position but also the ``wing" features in the off-center positions. 
The $\chi^{2}$ for the model is 1.8, larger than the inflow model.

In addition, we compare the position-velocity diagrams (PV diagrams) between the observational data, inflow model, and rotation model 
(Figure \ref{fig:pv}).
With the same angular resolution, the inflow model better demonstrates the observed discontinuity between the two velocity components.
The data shows an encounter of two velocity components at the position of zero offset. 
Such a feature is clearly seen in the inflow model but not in the rotation model at all.  
In summary, we suggest that the inflow model best describes the data and is the dominating mechanism for the kinematics.

\footnotetext[7]{To distinguish from 1D spherical infall, we use the term ``inflow" to describe gas infall on a 2D filament.}

\subsection{Small-scale Kinematics with CARMA}

\begin{figure*}
\begin{center}
\includegraphics[scale=0.85,angle=0]{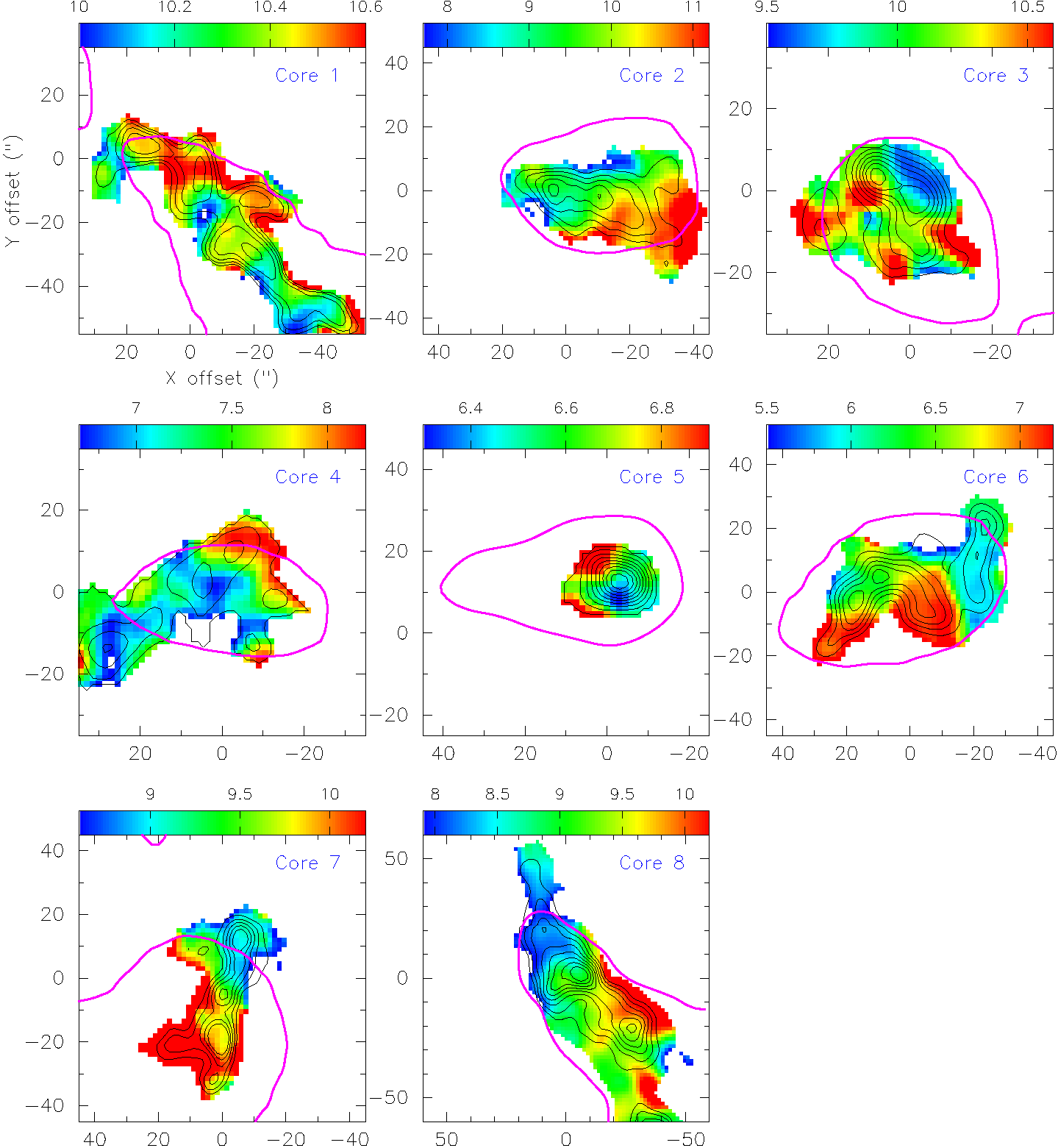}
\caption{First moment maps from the CARMA CS(2-1) data.  The magenta lines are the IRAM CS(2-1) cores.  The color indicates velocity in km s$^{-1}$.
The maps are masked based on the $5\sigma$ contours in the zeroth moment maps ($10\sigma$ for Core 6 and 8 to avoid too extended structures).
}
\label{fig:carma_mom1}
\end{center}
\end{figure*}

In the higher resolution CARMA maps, the kinematics of the cores is more complex (Figure \ref{fig:carma_mom1});
the interferometer resolves out large-scale structure in most cores or resolves multiple fragments with their
own velocity components.
Nonetheless, many of the cores, including Core 1, 2, 3, 4, and 8, show similar global behavior in the velocity patterns as the IRAM CS(2-1) results.

\section{Discussion}

\subsection{Fragmentation: Large and Small Scale}

Fragmentation in large-scale molecular clouds (few parsecs) to star-forming cores (0.1 pc) have been observed by previous observations.
At parsec-scales, molecular clouds have been extensively observed to fragment into clumps of sub-parsecs \citep[e.g.,][]{1998ApJ...502..296O,2007ApJ...665.1194I,2010A&A...518L.102A,2010A&A...520A..49S,2011A&A...533A..94H,2012ApJ...756...10L}.  
Several studies on the earliest stage of massive to intermediate star formation with high angular resolution (PdBI and SMA) have revealed fragmentation inside sub-parsec clumps to 0.1 pc cores \citep[e.g.,][]{peretto2006,zhang2009,2011A&A...530A.118P} at millimeter wavelengths.  
These studies have important implications to massive star and cluster formation.  

Limited by angular resolution and large distance to massive star-forming sites, 
the study of fragmentation inside 0.1 pc cores did not progress much until recently.  
With the angular resolution of 5\arcsec \ provided by CARMA and the
relatively small distance to Orion (414 pc) in this study, 
the CARMA observations reach a spatial resolution of $\sim 2000$ AU.  
Our observations revealed that multiple fragments are associated with each massive starless core of 0.1 pc (Sect.\ \ref{sect:carma_cs21}), 
suggesting that fragmentation continues to occur at 0.1 pc scales and 0.1 pc cores fragment to even smaller condensations.  
Recent observations with comparable angular resolutions have reported similar results of fragmentation inside 0.1 pc massive cores at millimeter wavelengths.
For example, \citet{2010A&A...524A..18B} observed a total of 23 fragments inside 5 massive dense cores in Cygnus X.
\citet{2011ApJ...735...64W} revealed 3 condensations of 0.01 pc inside two of the 0.1 pc cores in IRDC G28.34-P1.  
Among the 18 massive dense cores ($\sim$ 0.1 pc) that \citet{2013ApJ...762..120P} studied, 
$\ge$50\% showed $\ge4$ fragments and 30\% showed no signs for fragmentation.
Furthermore, using an even higher angular resolution of few hundred AUs and targeting nearby star-forming region Ophiuchus, 
\citet{2012ApJ...758L..25N} and \citet{2012ApJ...745..117B} unveiled the fragmentation inside low-mass 0.1 pc prestellar cores, 
suggesting a scenario beyond single collapse even for low-mass stars. 
Some of these condensations are prestellar in nature \citep{2010A&A...524A..18B,2012ApJ...758L..25N},
and some of them are protostellar evidenced by outflows \citep{2011ApJ...735...64W,2012ApJ...757...58N}.

\subsection{Mechanism for Fragmentation: Turbulence $+$ Magnetic Fields}
The results from the radiative transfer modeling (Sect.\ \ref{sect:radiative}) suggest a highly supersonic linewidth (0.8 km s$^{-1}$) for Core 2, implying that the environment is highly turbulent and turbulence is playing an important role in fragmentation.
Our modeling also showed that signs for fragmentation occur at the colliding point of the convergent flows.
This is broadly consistent with the ``turbulent fragmentation" scenario \citep[e.g.,][]{2005ApJ...620..786K,2002ApJ...576..870P}, 
in which density fluctuations are generated at small-scales (0.1 pc)
when large-scale shocks dissipate, which lead to star-forming cores.

Our results suggest the number of fragments associated with each massive starless core to be $\sim$ 3 - 5.  
This level of fragmentation is consistent with the recent studies of fragmentation inside massive dense cores in Cygnus-X \citep{2010A&A...524A..18B} and the two prestellar cores in $\rho$-Ophiuchus \citep{2012ApJ...758L..25N}. 
However, this number of fragments is not quite consistent with the prediction from several 
turbulent fragmentation models \citep[e.g.,][]{2004MNRAS.349..735B,2005MNRAS.360....2D,2005A&A...435..611J} 
since these models predict a much higher number of fragments.  
For example, \citet{2005MNRAS.360....2D} performed purely
hydrodynamical numerical simulations of a turbulent core of density structure $\rho \propto r^{-1.5}$ 
with a initial mass of 30 M$_{\sun}$ (comparable to that of our Core 2) and a radius of 0.06 pc.  
The study found that the core fragments into $\sim20$ objects.

The number of fragments can in principle be reduced by radiation feedback \citep{2007ApJ...656..959K} or magnetic fields \citep{2011A&A...528A..72H}.
The combination of the two effects work more efficiently in suppressing fragmentation \citep{2011ApJ...742L...9C,2012arXiv1211.3467M},
with the radiation feedback effectively suppressing the fragmentation in high-density regions (the center of the core) and 
magnetic fields effectively suppressing the fragmentation in
low-density regions (the outer part of the core).
However, since the cores we studied are starless, the radiation 
feedback is expected to be weak.   
On the other hand, large-scale magnetic fields are observed in the
main region of OMC-1 that is close to our cores \citep[e.g.,][]{2004ApJ...604..717H}, supporting the argument that magnetic fields may play a role in the fragmentation process.  
Measurements of magnetic fields on small-scales at the position of the starless cores are needed to further determine the precise role of magnetic fields.

\subsection{Role of Supersonic Converging Flows}
\label{sect:flow}

We obtained a dynamical velocity pattern in the supersonic converging flows (Sect.\ \ref{sect:radiative}) associated with Core 2 from the radiative transfer modeling. 
Other cores including Core 1, 4 and 8 all showed similar spectral features as Core 2 (see Sect \ref{sect:spectra_cs21}) which can be explained by converging flows.
Only a few studies have detected the sign for such supersonic convergent flows at 0.1 pc scale \citep{2011A&A...527A.135C,2011ApJ...740L...5C}. 
The origin of the supersonic flows is difficult to identify;
however, it is natural to speculate the origin being the large-scale flows at few parsecs scale since the prominent large-scale filamentary structures may be due to large-scale turbulent flows \citep[e.g.,][]{2004RvMP...76..125M}.

Is the converging flow dynamically important in the star formation process?
We calculated the flow crossing time as: $ t_{cross} = R (0.05 pc) / v_{inf} (1.5 km/s) = 3.0 \times 10^4$ yr, 
where $v_{inf}$ is the inflow velocity derived from the radiative transfer modeling.
The free-fall time is $t_{ff} = \sqrt{\frac{3\pi}{32G\rho}} = 1.5\times 10^4$ yr, 
where we use the derived central density for $\rho$ in the calculation
and therefore the derived free-fall time is an upper limit. 
The flow-crossing time is the timescale that the flows at 0.1 pc scale bring material down to the center, 
and the free-fall time is the timescale for the material to collapse gravitationally.   
The flow crossing time is comparable to the free-fall time, suggesting that the flow is dynamically important 
in forming the density condensations.
However, the crossing time is slightly larger than the free-fall time, 
suggesting that gravity takes over at small-scale in driving the dynamical process of star formation.
We posit that large-scale flows initiate the density condensations, and 
gravity becomes dominant at small-scales which enhances the converging flows.

We also estimated the mass inflow rate along an filament: $\dot{M}_{inf}=2\times \pi R^{2}_{fil} \times n_{mean} \times \mu \times m \times v_{inf} = 2.35 \times 10^{-3}$ M$_{\sun}$ yr$^{-1}$, 
where $R_{fil} = 0.025$ pc is the radius of the cylinder (estimated from the model; see Fig.\ \ref{fig:model}), $n_{mean}$ is the mean number density of the cylinder, $\mu$ is the mean molecular weight, and $m$ is the mass of hydrogen.
The LTE mass of Core 2 is estimated to be $36.6$ M$_{\sun}$ (Sect.\ \ref{tbl:coldensity}), 
and therefore the formation timescale for Core 2 is $\sim 1.5 \times 10^{4}$ yrs.  
The inflow velocity and mass inflow rate appear to be large compared to low-mass stars which typically have infall velocities $\sim$ 0.1 km s$^{-1}$ \citep[e.g.,][]{2004ApJS..153..523L}.
However, higher inflow velocities and mass inflow rates are not surprising for massive star-forming regions.
For example, several high infrared extinction clouds with massive star formation in \citet{2013A&A...549A...5R} have the spherical infall velocities in the order of 0.3 - 7 km s$^{-1}$ and mass infall rates on the order of 1.4 - 22.0$\times 10^{-3}$ M$_{\sun}$ yr$^{-1}$.
\citet{peretto2006} reported a similar mass inflow rate in a protocluster NGC 2264-C with intermediate- to high- mass star formation.
The study found a mass inflow velocity of 1.3 km s$^{-1}$ along a cylinder-like, filamentary structure on a spatial scale of 0.5 pc. 
We are probably seeing the continuation of the 0.5 pc - scale flow down to the scale of 0.1 pc.    
However, the underlying physical reason for such highly supersonic velocities is still unclear.  

\subsection{Implications for Massive Star and Cluster Formation}

Our result does not fully support either the turbulent core scenario or the competitive accretion scenario.  
The discovery of fragments in our study makes it harder to form
  massive stars in these cores via the turbulent core model proposed by
  \citet{2003ApJ...585..850M}, since the core mass will eventually go
  to a number of objects rather than a single star. 
%
\citet{2003ApJ...585..850M} suggests a minimum mass accretion rate of $10^{-3}$ M$_{\sun}$ yr$^{-1}$ to 
overcome the radiation pressure and form a massive star.
By this criterion, our Core 2 has a high enough inflow rate to 
form massive stars; however, the inflow may not feed just one single
object since the core contains multiple fragments.  
\citet{2008Natur.451.1082K} suggests that a minimum column
  density of 1 g cm$^{-2}$ can avoid fragmentation through radiative
  feedback. We suggest that this is a necessary but not sufficient
  condition for massive star formation: all cores in this study have
  column densities larger than that threshold (see Table
  \ref{tbl:coldensity}), and yet they have already fragmented before
  the radiative feedback kicks in.

Our result is broadly consistent with a scenario of turbulent
fragmentation, with the number of fragments perhaps reduced by magnetic fields.  
However, while it is possible that the fragmentation continues to
occur during the later stages of evolution and/or future massive stars
could form via competitive accretion
\citep{2004MNRAS.349..735B,2006MNRAS.370..488B}, our observations
highlighted a feature that is not present in the standard competitive
accretion scenario: rapid converging flows along dense filaments that
feed matter into the central region. This feature is similar in spirit
to the model proposed by \citet{2010ApJ...709...27W} where the
mass accretion rate onto a massive star is set mainly by the 
large-scale converging or collapsing flow, rather than the
gravitational pull of the star itself.   
%
%

As each fragment has the potential to collapse individually and form protostars, 
it is suggestive that we are witnessing the formation of clusters at the very early stages and multiplicity occurs already in the prestellar phase.  
Given a 30\% core formation efficiency \citep{2010A&A...524A..18B} for a 30 M$_{\sun}$ core (Sect.\ \ref{tbl:coldensity}),
each individual fragment inside a core will be forming low- to intermediate- mass stars.  
Therefore, we suggest that these cores are in the dynamical state of forming low- to intermediate- mass protoclusters \citep[e.g.,][]{2011MNRAS.415.2790L}.

Although our study is unique in observing massive starless cores at a distance $\le500$ pc with high spatial resolution, 
the number of observed fragments may increase with higher resolution and more sensitivity.
Follow-up studies of dust continuum with higher resolution 
are necessary to compliment this study and accurately constrain the properties of these cores, including the masses and the dynamical states of these cores.

\section{Conclusion}

We observed nine starless cores in the Orion-A North region with the IRAM 30-m telescope and eight cores out of the nine cores with CARMA using CS(2-1). Our main conclusions are as follows: 
\begin{enumerate}
  \item 
The IRAM 30-m observations showed no detection of N$_{2}$D$^{+}$(2-1) for all the nine cores, and the CARMA observations showed N$_{2}$H$^{+}$(1-0) for only two cores (Core 4 and Core 8).  
As CS(2-1) is regarded as an ``early-time tracer" and N-bearing species (N$_{2}$D$^{+}$(2-1), N$_{2}$H$^{+}$(1-0)) as ``late-time tracers", 
this result suggests that most of our cores are at the very early stage of star formation.  
  \item 
The CS(2-1) observations with the IRAM 30-m telescope showed that majority of the starless cores are single-peaked, and the morphologies traced by CS(2-1), C$^{34}$S(2-1) and CS(3-2) are mostly consistent with each other. 
The column densities estimated from CS(2-1) range from $7-42 \times 10^{23}$ cm$^{-2}$ and the LTE masses range from 20 M$_{\sun}$ to 154 M$_{\sun}$. 
  \item
The comparison between the CARMA CS(2-1) data, the IRAM CS(2-1) data,
JCMT 850 $\um$ dust continuum, Herschel 500 $\um$ data shows that 
gas structures probed by CS(2-1) are forming along small-scale filamentary structures traced by dust continuum, suggesting the importance of 
filamentary structures to star formation even at small scales.
\item 
The CARMA CS(2-1) observations show fragmentation inside all the cores except for Core 5.
The number of fragments associated with each core ranges from 3 to 5.
  \item
Five cores showed obvious velocity gradients across the cores in the IRAM CS(2-1) data.  
We performed a two-dimensional fitting to the velocity gradients by assuming the velocity gradients are linear. 
The fitting results showed that the velocity gradients range from 1.7 - 14.3 km s$^{-1}$ pc$^{-1}$.
  \item
Four cores (Core 1, Core 2, Core 4, Core 8) showed two common features in their spectra along the direction of the velocity gradients.  
First, the velocity peak changes from blue-shifted to red-shifted across the cores.
Second, the intensity of the blue peak is always stronger than the red peak. 
  \item 
We propose a model of a cylindrically symmetric filament with
converging inflows from the two sides toward the center to explain the two spectral features.
We modeled Core 2 with this proposed kinematic model with the radiative transfer code LIME \citep{2010A&A...523A..25B}, and verified that the kinematic model successfully explains the two features.
The best-fit gives a constant supersonic speed of 1.5 km s$^{-1}$ for the flow velocity and a supersonic linewidth of 0.8 km s$^{-1}$.
A mass inflow rate of $2.35\times 10^{-3}$ M$_{\sun}$ yr$^{-1}$ is inferred from the inflow velocity.
  \item
The supersonic linewidth from the modeling suggests that the core environment is highly turbulent and the fragmentation revealed by the CARMA observations may be due to turbulent fragmentation. 
However, the number of fragments is much less than the predictions
from turbulent fragmentation models
\citep[e.g.,][]{2005MNRAS.360....2D}, indicating that magnetic fields
may be playing an important role in reducing the level of fragmentation \citep{2011A&A...528A..72H}.    
  \item 
The small-scale converging flow is dynamically important to the formation of the cores and their substructures.
We suggest that large-scale flows initiate the density condensations, 
and gravity becomes dominant at small-scales which enhances the converging flows.
Due to the high mass inflow rate, each fragment is likely to collapse individually and form seeds for future protoclusters. 
Given a core formation efficiency of 30\%, we suggest that these cores are in the dynamical state of forming low- to intermediate- mass protoclusters.
  \item
The fragmentation observed in our cores makes massive star
  formation via the turbulent core model proposed by
  \citet{2003ApJ...585..850M} more difficult. Our result does not
  fully support the standard competitive accretion model either, since
  it does not account for our inferred rapid inflow along filaments,
  which may be an important way of feeding massive protostars.  
\end{enumerate}

\section{Acknowledgement}
We thank the anonymous referee for valuable comments that improved this paper. We acknowledge support from the Laboratory for Astronomical Imaging at the University of Illinois. Support for CARMA construction was derived from the states of Illinois, California, and Maryland; the Gordon and Betty Moore Foundation; the Eileen and Kenneth Norris Foundation; Caltech Associates; and the National Science Foundation. Ongoing CARMA development and operations are supported by the National Science Foundation, and by the CARMA partner universities.  We also acknowledge the support from the National Radio Astronomy Observatory.  The National Radio Astronomy Observatory is a facility of the National Science Foundation operated under cooperative agreement by Associated Universities, Inc.  Z. Li acknowledges the support in part by NASA grant NNX10AH30G.

\bibliographystyle{apj} 
\bibliography{paper.bib} 
\clearpage

\end{document}